\documentclass[aps,pre,showpacs,superscriptaddress,longbibliography,twocolumn,floatfix]{revtex4-1}
\usepackage{graphicx}
\usepackage{amsmath}
\usepackage{amssymb}
\usepackage{mathtools}
\usepackage{makeidx}
\usepackage{amsfonts}
\usepackage{bm}
\usepackage[colorlinks]{hyperref}
\begin{document}

\title{Replica Symmetry Broken States of some Glass Models}
\author{J.\ Yeo}
\affiliation{Department of Physics, Konkuk University, Seoul 05029, Korea}
\author{M.\ A.\ Moore}
\affiliation{Department of Physics and Astronomy, University of Manchester,
Manchester M13 9PL, United Kingdom.}

\date{\today}

\begin{abstract}
We have studied in detail the $M$-$p$ balanced spin glass model, especially the case $p=4$. These types of model have relevance to structural glasses. The models possess two kinds of broken replica states; those with one-step replica symmetry breaking (1RSB) and those with full replica symmetry breaking (FRSB). To determine which arises requires studying the Landau expansion to quintic order. There are 9 quintic order coefficients, and 5 quartic order coefficients, whose values we  determine for this model. We show that it is only for $2 \leq M < 2.4714 \cdots$ that the transition at mean-field level is to a state with FRSB, while for larger $M$ values there is either a continuous transition to a state with 1RSB (when $ M \leq 3$) or a discontinuous transition  for $M > 3$. The Gardner transition from a 1RSB state at low temperatures to a state with FRSB also requires the Landau expansion to be taken to quintic order. Our result for the form of FRSB in the Gardner phase is similar to that found when $2 \leq M < 2.4714\cdots$, but differs from that given in the early paper of Gross et al.~[\href{https://doi.org/10.1103/PhysRevLett.55.304}{Phys.\ Rev.\ Lett.\ {\bf 55}, 304 (1985)}]. Finally we discuss the effects of fluctuations on our mean-field solutions using the scheme of H\"{o}ller and Read [\href{https://doi.org/10.1103/PhysRevE.101.042114}{Phys.\ Rev.\ E {\bf 101}, 042114 (2020)}] and argue that such fluctuations will remove both the continuous 1RSB transition and discontinuous 1RSB transitions when $8 >d \geq 6$  leaving just the FRSB continuous transition. We suggest values for $M$ and $p$ which might be used in simulations to confirm whether fluctuation corrections do indeed remove the 1RSB transitions.

\end{abstract}



\maketitle

\section{Introduction}
\label{sec:intro}

Spin models of the $p$-spin or Potts glass variety \cite{gross:85,gardner:85} played an important role in the development of one of the current theories of structural glasses, the Random First Order Transition (RFOT) picture \cite{kirkpatrick:89,kirkpatrick:15,lubchenko:07,cavagna:09,biroli:09}. These models have been primarily studied in the infinite dimensionality limit, which is equivalent to mean-field theory. Of course what is really wanted is an understanding of what happens in the physical realm of two and three dimensions, and for these dimensions simulations \cite{franz:99,campellone:98} of models of the type studied in this paper have revealed that they behave completely differently from what is predicted by the mean-field calculations. In particular in the simulations there is no sign of the random first-order transition which is one of the central features of RFOT theory.  Below the ideal glass transition there is supposed to exist the ideal glass state, a state of low configurational entropy but with a high stability due to the assumed paucity of glass states. This state in replica language has one-step replica symmetry breaking (1RSB). The transition temperature to this state is identified  as the Kauzmann temperature in RFOT theory, which is the temperature at which the entropy of the glass state becomes equal to that of  the crystalline state \cite{kauzmann:48}. While a discontinuous  transition was not seen in the simulations, evidence was found for the existence of long correlation lengths, which is also the behavior found in real-space renormalization group (RG) calculations \cite{yeoB:12, yeo:13} of $p$-spin models in three dimensions.

That simulations in three dimensions lead to a picture quite different to that which arises from mean-field calculations has largely been ignored: Work has continued apace using the large $d$ limit and mean-field techniques. We have therefore begun a program of trying to understand why the mean-field picture  does not extend to three dimensions \cite{yeo2020possible}. For one particular $p$-spin model, the $M$-$p$ spin glass model with $p=6$, we were able to give an argument that the 1RSB state of that model was unstable in any finite dimension due to the excitation of droplets of flipped spins whose interface free energy are very small \cite{moore:06b}. That argument is specific to glass models with a particular form of time reversal symmetry which gives rise to a field theory in which the cubic term $w_2$ is zero (see Eq.~(\ref{quinticLG})). Unfortunately the generic field theories thought relevant to glasses have $w_2$ non-zero and it is these which we study in this paper. Most of our work will be focussed on the case of $p=4$. The 1RSB phase for $p=6$ spin glasses is destroyed by non-perturbative droplet excitations. For generic glass models with $w_2$ non-zero, we can only find perturbative arguments. They are strong enough to lead us to the conclusion that the continuous phase transition to a state with 1RSB will not exist for dimensions $d$ less than $8$ and will be replaced by a continuous transition to a state with FRSB. We shall suggest that fluctuation corrections to the coupling terms in Eq.~(\ref{quinticLG}) might also drive the system away from having a discontinuous transition to a 1RSB state  to  a continuous transition to a state  with full replica symmetry breaking (FRSB), but we do not know whether the fluctuation corrections are large enough to bring that about. We suspect that this question will only be resolved by simulations and values of $p$ and $M$ which might be appropriate for such simulations are suggested in Sec.~\ref{sec:discussion}.

Our procedure is based upon the old idea \cite{rudnick:76} of using the renormalization group recursion relations for the coupling constants of the field theory to map the coefficients of the critical field theory into a region where the correlation lengths are small and Landau theory (i.e.\ mean-field theory) with small fluctuation corrections can be  employed. This program has also been used by H\"oller and Read \cite{holler:20} on the problem of the de Almeida-Thouless transition of the Ising spin glass in a field \cite{almeida:78}. It has a field theory identical to that of the $M$-$p$-spin glass models discussed in this paper, i.e.\ that of Eq.~(\ref{quinticLG}), but with different numerical values for the coefficients. (To discuss finite dimensions a gradient term of the form $\int d^d r \sum_{a,b} (\nabla q_{ab} (r))^2$ would need to be included in Eq.~(\ref{quinticLG}).) The program therefore requires us to understand in detail the stationary solutions i.e.\ mean-field solutions of Eq.~(\ref{quinticLG}), and the bulk  of this paper is devoted to this task. Because H\"{o}ller and Read discussed the RG aspects of the calculations in great detail, we shall treat those briefly, just focussing on the implications  of numerical studies which were carried out  after their paper was written \cite{aguilar:23}.

In Sec.~\ref{sec:balancedmodel} we introduce the balanced $M$-$p$ models and the replica procedure which was used to average their free energy over disorder. The balanced $M$-$p$ spin models are very convenient to study with simulations as they are readily extended to finite dimensions on a $d$-dimensional lattice. When this is done the resulting field theory acquires the  already mentioned gradient squared term. One of the attractions of the balanced version of these models is the absence of \lq\lq hard modes", which are just usually cast aside (as in  the paper of Caltagirone et al.\ \cite{caltagirone:11}), but this leaves the subsequent calculations of uncertain accuracy. We shall focus on the case $p=4$ and regard the number of types of Ising spins $M$ as a variable which can take non-integer values. The simulations of Campellone et al.\ \cite{campellone:98} which failed to find a discontinuous 1RSB transition were in fact done for a closely related model with $p=4$ and $M=4$ in three dimensions. At cubic order there are two coupling constants, $w_1$ and $w_2$, at quartic order, there are five coupling constants, $y_1,\cdots, y_5$ and at quintic order, there are nine coupling constants, $z_1,\cdots,z_9$. The quadratic term $\tau$ vanishes as usual at the mean-field transition temperature $T_c$ and is negative when $T < T_c$.  We calculate the \lq\lq bare" value of all these coefficients in Appendix A for the case $p=4$. Fluctuation corrections will modify the bare values. In studying the model at non-integer values of $M$ we are anticipating that the fluctuation corrections can modify the bare coefficients. Studying the field theory of Eq.~(\ref{quinticLG}) for general values of the coefficients would be a good idea, but there are so many of these coefficients that we have limited our study to those values which can be reached by varying $M$ in the bare values. In Sec.~\ref{sec:discussion} we discuss what we believe will be the likely consequences of fluctuation effects on the coupling constants.
 
In Sec.~\ref{sec:replicasymmetric} we determine the free energy of the system in the high-temperature or paramagnetic phase where the order parameter $q_{ab}$ is independent of $a$ and $b$, that is, replica symmetric. At mean-field level $q_{ab}=0$, (but fluctuation corrections would leave it replica symmetric but non-zero). If the transition is continuous, so that $q_{ab}$ is small just below the transition, then the expansion of the Landau-Ginzburg free energy functional in powers of $q_{ab}$ should be useful and we give its form in Sec.~\ref{sec:landauexpansion}. Most workers have stopped at the quartic terms, but we have continued up to the quintic terms. This is necessary for two reasons. The difference in free energy between the 1RSB free energy and the FRSB free energy is of $O(\tau^5)$, (see for example, Ref.~\cite{aspelmeier:08}). Thus one needs to worry about the quintic terms  when working out whether the state which forms at the continuous transition is of 1RSB type or is of FRSB type. Fortunately, we can show that the borderline value of $M$, $M^{**} \approx 2.47140$ between these types is not dependent on the quintic terms. (For $2 \leq  M < M^{**}$ the continuous transition is to a state with FRSB, while for $M^{**} < M < 3$, the continuous transition is to a state with 1RSB.) The second reason relates to studies of the Gardner transition \cite{gross:85,gardner:85}. The Gardner transition is the transition from a state with 1RSB to a state with FRSB as the temperature is lowered. Right from the beginning it was realized that the  quintic terms are needed for its study \cite{gross:85}. We shall find though that our actual FRSB solution is quite different to that of Ref. \cite{gross:85}. This is discussed in Sec. \ref{sec:frsb}.

A feature of the FRSB solutions is a singularity first noticed by Goldbart and Elderfield \cite{goldbart:85}. They found that the FRSB solution for $q(x)$ at quartic level could have an  unphysical singularity in the interval $0 < x <1$ which would imply that the probability of two states having an overlap $q$ would be negative, which is impossible. This problem was studied in some detail by Jani\v{s} and colleagues using a non-standard approach to replica symmetry breaking \cite{janis:13}. We find in Sec. \ref{sec:frsb} that the singularity at quartic level in fact determines the value of $M^{**}$ and that one avoids the singularity at $M > M^{**}$ by simply being in the state with 1RSB. At the Gardner transition the quintic terms remove the  quartic level singularities. However, similar singularities are to be found also at quintic level.  Right at the Gardner transition temperature $T_G$ , just where the free energies of the FRSB state and the 1RSB state are equal, the Goldbart-Elderfield singularity is at the lower breakpoint $x_1$. This causes the derivative of $q(x)$ at $x=x_1$ to be infinite. However for temperatures $T$ less than $T_G$, the singularity is below $x_1$ and the derivative stays finite. 

In Sec.~\ref{sec:1rsb} we derive the free energy at mean-field level for the 1RSB state. For $M >3$, when $w_2/w_1>1$, the transition from the high temperature normal phase to a state with 1RSB is a discontinuous transition which takes place at a transition temperature above $T_c$. We suspect that this behavior would be seen for all values of $M> 3$. However, if one truncates the free energy to quartic level terms, as is commonly done, the 1RSB state only exists in the interval $3< M\lesssim 6.64$. With the inclusion of the quintic terms, the 1RSB forms at a discontinuous transition when  $14.41 \gtrsim M \gtrsim 3.98$   and $3.27 \gtrsim M > 3$. Thus with the quintic form the 1RSB state persists up to larger values of $M$. We believe that if all terms were kept then the discontinuous transition to the  1RSB state would exist for all $M > 3$. In Sec.~\ref{sec:largem} we describe the simplifications which arise in the large $M$ limit. Truncation leads to spurious features as the Landau expansion cannot be expected to be accurate when $q_{ab}$ is not small. Another spurious feature of truncation is the apparent phase transition at low temperatures from the 1RSB state to the replica symmetric state with $q_{ab}$ non-zero. In the large $M$ limit we can solve without truncation and such a transition does not arise (see Sec.~\ref{sec:1rsb}). 

The form of the FRSB solutions at both quartic and quintic level, together with the Gardner transitions, is in Sec.~\ref{sec:frsb}. In Sec.~\ref{sec:discussion} we discuss how fluctuation corrections to the coupling constants used in the mean-field solution will change the continuous 1RSB transition into the continuous FRSB solution, using extensions of the approach of H\"{o}ller and Read \cite{holler:20}.  We suspect that the discontinuous 1RSB transition might also suffer the same fate, based on the results of simulations in low dimensions \cite{franz:99,campellone:98}, but we cannot support this possibility with  analytical arguments.  We finally conclude with suggestions of the kinds of model which could be studied numerically  to resolve these issues, and also to resolve the question of whether the FRSB state can exist for dimensions $d < 6$.

\section{The balanced $M$-$p$ model in the fully connected limit}
\label{sec:balancedmodel}
In this section, we study the $M$-$p$ spin glass model in the fully connected limit,
where one has $M$ different types of Ising spins, $S_i(x)$, $i=1,2,\cdots,M$ at each site $x$
coupled with spins on other sites via $p$-body interactions. 
Here we focus on the so-called balanced model introduced in Ref.~\cite{yeo2020possible} for even $p$,
where only the coupling between two sets of $p/2$ spins on two different sites is considered. 
It amounts to considering only the soft mode in a more general $M$-$p$ model, where all the couplings
between $k$ spins and $p-k$ spins are included for $k=1,2,\cdots ,p-1$.

In this paper, we focus on the $p=4$ case. For $p=4$, the balanced model is given by four-spin interactions between a pair of two spins
on two different sites. Each site has ${M \choose 2}$ different two-spin combinations.
Therefore, for given pair of sites, there are ${M\choose 2}^2$ terms in the Hamiltonian. 
The Hamiltonian is given by
\begin{align}
H=-\frac 1 2 \sum_{x\neq y}&\Big[ \sum_{i_1<i_2}^M\sum_{j_1<j_2}^M J^{(i_1,i_2),(j_1,j_2)}_{x,y} \nonumber \\
&\times S_{i_1}(x)S_{i_2}(x)
S_{j_1}(y)S_{j_2}(y)
\Big] ,
\end{align}
where each $J^{(i_1,i_2),(j_1,j_2)}_{x,y}$ is drawn from the Gaussian distribution with zero mean and the variance
\begin{equation}
\frac{J^2}{NM^{p-1}}=\frac {J^2} {NM^3}.
\label{variance}
\end{equation}
We will set $J=1$ for convenience. 
After neglecting the terms of subleading order in $N$, we can write
the replicated partition function averaged over the disorder as
\begin{align}
\overline{Z^n}
=&\mathrm{Tr} \exp\Big[ \frac{\beta^2}{4NM^3} \\ 
&\times \sum^n_{a,b} 
\Big\{ \sum^N_x 
\sum_{i_1<i_2}^M S^a_{i_1}(x)S^a_{i_2}(x)S^b_{i_1}(x)S^b_{i_2}(x) \Big\}^2\Big] .  \nonumber
\end{align}
The diagonal terms ($a=b$) in the replica indices give a factor $\exp[nN\beta^2 C]$ where
\begin{align}
C=\frac{1}{4M^3}{M\choose 2}^2= \frac{(M-1)^2}{16 M}.
\label{c_bal}
\end{align}
For $a\neq b$, following the convention used in Ref.~\cite{caltagirone:11}, we introduce the delta functions
enforcing
\begin{align}
q_{ab}=\frac 1 {NM^2} \sum^N_x \sum_{i_1<i_2}^M S^a_{i_1}(x)S^a_{i_2}(x)S^b_{i_1}(x)S^b_{i_2}(x) 
\end{align}
in the replicated partition function. Using the integral representation of the delta function, we can write
\begin{align}
\overline{Z^n}
=e^{nN\beta^2 C}\int\prod_{a<b} dq_{ab}  d\mu_{ab}  \;\exp[-NG(
\underline{q},\underline{\mu})], \label{zn}
\end{align}
where
\begin{equation}
G(
\underline{q},\underline{\mu})=- \frac M 4 \beta^2\sum_{a\neq b}q^2_{ab}
+\frac M 2\sum_{a\neq b}\mu_{ab}q_{ab} - \ln L(\underline{\mu})
\label{G}
\end{equation}
and 
\begin{align}
\label{L}
L(\underline{\mu})=\underset{\{S_i^a\}}{\mathrm{Tr}} \exp\Big[   \frac 1 {2M} \sum_{a\neq b} \mu_{ab}\sum^M_{i<j}
 S^a_i S^a_j  S^b_i S^b_j \Big].
\end{align}
In the large-$N$ limit, the integral is dominated by the saddle points which are determined by
\begin{align}
\mu_{ab}=\beta^2 q_{ab} \label{lq}
\end{align}
and 
\begin{align}
q_{ab}=\frac 1 {M^2} \left\langle  \sum^M_{i<j} S^a_i S^a_j  S^b_i S^b_j \right\rangle_L ,
\end{align}
where $\langle\cdots\rangle_L$ is evaluated with respect to $L$ in Eq.~(\ref{L}).
The free energy $F$ is then given by
\begin{align}
\frac{\beta F}{N}&=-\frac 1 N \lim_{n\to 0}\frac 1 n \ln \overline{Z^n} 
=-C\beta^2 +\lim_{n\to 0}\frac 1 n G(\underline{q},\underline{\mu}).
\label{free1}
\end{align}

\subsection{Replica Symmetric Solution}
\label{sec:replicasymmetric}
We first look for the saddle point solutions in the replica symmetric (RS) form $q_{ab}=q$ and 
$\mu_{ab}=\mu$ for all $a\neq b$.
We have
\begin{align}
\lim_{n\to 0}\frac 1 n G(q,\mu)=\frac M 4 \beta^2 q^2  - \frac M 2 \mu q  - \lim_{n\to 0}\frac 1 n \ln L(\mu).
\end{align}
Using 
\begin{align}
\sum_{a\neq b} 
 S^a_i S^a_j  S^b_i S^b_j =\left(\sum_a  S^a_i S^a_j \right)^2 -n 
\end{align}
in Eq.~(\ref{L}) and
the Hubbard-Stratonivich transformation on the first term, we can rewrite Eq.~(\ref{L}) as
\begin{align}
L(\mu)=&e^{-n\mu (K/2M)}\;\underset{\{S_i^a\}}{\mathrm{Tr}}\;
 \int D^K \bm{y}\nonumber \\
 &\times \exp\left[ \sqrt{\frac{\mu}M}\sum_a \sum_{i<j}^M y_{(i,j)} S^a_i S^a_j \right] ,
\end{align}
where 
\begin{equation}
K\equiv {M\choose 2}
\end{equation}
and the integral over the $K$-dimensional vector $\bm{y}=(y_1,y_2,\cdots,y_K)\equiv(y_{(1,2)},y_{(1,3)},\cdots,y_{(M-1,M)})$
is defined as
\begin{equation}
\int D^K\bm{y} \equiv \prod_{\alpha=1}^K\left( \int_{-\infty}^\infty  \frac{dy_\alpha}{\sqrt{2\pi}}e^{-y^2_\alpha/2} \right).
\end{equation}
We therefore have
\begin{align}
 \lim_{n\to 0}\frac 1 n \ln L(\mu)= -\frac K {2M} \mu
    +M\ln 2 +\int  D^K\bm{y} \; \ln 
 \zeta(\bm{y},\mu),
\end{align}
where 
\begin{align}
\zeta(\bm{y},\mu)\equiv \frac{1}{2^M}\underset{\{S_i\}}{\mathrm{Tr}}\;  \exp\left[ \sqrt{\frac{\mu}{M}}\bm{y}\cdot\bm{\Psi}
\right]
\label{xidef}
\end{align}
with the $K$-dimensional vector
$\bm{\Psi}=(\Psi_1,\Psi_2,\cdots,\Psi_K)=(S_1 S_2, S_1 S_3,\cdots, S_{M-1}S_{M})$.
The RS free energy is then given by
\begin{align}
\frac{\beta F_{\rm RS}}{N}=& -C \beta^2 +\frac M 4 \beta^2 q^2  - \frac M 2 \mu q +\frac K {2M} \mu \nonumber \\
&-M\ln 2 - \int D^K\bm{y}\; \ln \zeta(\bm{y},\mu).
\end{align}

By varying the free energy with respect to $q$ and $\mu$, respectively, we have saddle point equations,
\begin{align}
\mu=\beta^2 q
\end{align}
and
\begin{align}
q=&\frac 1 {M^2} \int D^K \bm {y}\; \frac 1 {\zeta^2(\bm{y},\mu)}\nonumber \\
&\times \sum_{\alpha=1}^K \left\{ \frac 1 {2^M} \underset{\{S_i\}}{\mathrm{Tr}}\; \Psi_\alpha 
\exp\left[\sqrt{\frac{\mu}{M}}\bm{y}\cdot\bm{\Psi}\right] \right\}^2 .
\end{align}
At high temperatures, the RS solutions are given by $q=\mu=0$. In that case,
$\zeta=1$ and the corresponding free energy is 
\begin{equation}
\frac{\beta F_{\rm RS}}{N}=-C\beta^2 -M\ln 2. 
\label{free_highT}
\end{equation}
The entropy $S=-\partial F/\partial T$ for this phase is
\begin{equation}
\frac{S_{\rm RS}}{N}=-C\beta^2+M\ln 2.
\end{equation}
This becomes negative below 
\begin{equation}
T_*=\sqrt{\frac {C}{M\ln 2}}=\frac{M-1}{4M\sqrt{\ln 2}}.
\label{tstarm}
\end{equation}
Some values of $T_*$ are $T_*$=0.20019 for $M=3$, 0.22521 for $M=4$, 0.25023 for $M=6$ and 0.25738 for $M=7$. It keeps increasing with $M$ and 
approaches 0.30028 in the $M\to\infty$ limit. 

\subsection{Landau Expansion of Free Energy}
\label{sec:landauexpansion}
In order to study a possible continuous transition, 
we expand the free energy, Eq.~(\ref{free1}) for small values of the order parameter. 
We first expand Eq.~(\ref{L}) to $O(\mu^5)$ and take the trace over the spins.
The detailed steps are given in Appendix \ref{app:exp}.
Now using Eqs.~(\ref{G}), (\ref{lq}) and (\ref{free1}), we can write the free energy as
\begin{align}
\frac{\beta F}{N} &=   -C\beta^2  -M\ln 2 
+\lim_{n\to 0}\frac 1 n \Big[ \tau
\sum_{a, b} q^2_{ab} \label{quinticLG} \\ 
& -w_1 \sum_{a,b,c}q_{ab}q_{bc}q_{ca} 
- w_2 \sum_{a, b}q^3_{ab} 
-y_1 \sum_{a, b} q^4_{ab} \nonumber \\
&-y_2 \sum_{a,b,c}q^2_{ab}q^2_{bc} 
-y_3\sum_{a,b,c}q^2_{ab}q_{bc}q_{ca} 
-y_5 \sum_{a,b,c,d}q_{ab}q_{bc}q_{cd}q_{da} \nonumber \\
&-  z_1  \sum_{a,b} q^5_{ab} 
 -z_2 \sum_{a,b,c} q^3_{ab}q^2_{bc}
 - z_3  \sum_{a,b,c} q^3_{ab}q_{bc}q_{ca} \nonumber \\
 &- z_4 \sum_{a,b,c} q^2_{ab}q^2_{bc}q_{ca}
 -z_5 \sum_{a,b,c,d} q^2_{ab}q_{bc}q_{cd}q_{da} \nonumber \\
& -  z_6  \sum_{a,b,c,d} q^2_{ab}q_{bc}q_{cd}q_{db} 
 - z_7 \sum_{a,b,c,d} q^2_{ab}q_{bc}q^2_{cd} \nonumber \\
&    -z_8 \sum_{a,b,c,d} q_{ab}q_{bc}q_{cd}q_{da}q_{ac} 
  -z_9 \sum_{a,b,c,d,e} q_{ab}q_{bc}q_{cd}q_{de}q_{ea} \Big], \nonumber
\end{align}
where $q_{aa}=0$, $q_{ab}=q_{ba}$, and all the sums over replica indices are without any restriction.

The coefficient of the quadratic term  is given by
\begin{align}
\tau=\frac M 4 \beta^2 \left(1- \frac K {M^3} \beta^2\right)=\frac M 4 \beta^4 \left(T^2-T^2_c\right),
\end{align}
where 
\begin{equation}
T_c\equiv\sqrt{\frac K {M^{3}}}  =  \frac{1}{M}\sqrt{\frac{M-1}{2}}.
\label{Tc}
\end{equation}
This expression coincides with Eq.~(27) of Ref.~\cite{caltagirone:11}.
Some values of $T_c$ are 0.33333 for $M=3$, 0.30619 for $M=4$, 0.26352 for $M=6$ and 0.24744 for $M=7$.
Note that $T_c$ decreases with $M$ and becomes zero in the $M\to\infty$ limit.
Note also that $T_c>T_*$ for $M=2,3,\cdots,6$ and $T_*>T_c$ for $M\ge 7$.

The coefficients of the cubic terms are given by
\begin{align}
w_1 = \frac {\beta^6 K} {6M^3},~~w_2 =\frac {\beta^6 K} {6M^3}  (M-2).
\label{w12}
\end{align}
The quartic and quintic coefficients are given in Appendix \ref{app:exp} as functions of $M$.
It is known \cite{gross:85,caltagirone:11} that if the ratio of the cubic terms $w_2/w_1$, which in our model is equal to $M-2$, is greater than one, a discontinuous transition
to the one-step replica symmetry breaking phase (1RSB) occurs.
When $M=2$, our model reduces to the Ising spin glass
and we can check that the cubic and quartic coefficients coincide with those for the Ising spin glass
except for the multiplicity factor of $2^3$ for $w_i$ and $2^4$ for $y_i$.

\subsection{The 1RSB Solution}
\label{sec:1rsb}

We now consider the case where $q_{ab}$ and $\mu_{ab}$ take the one step replica symmetry breaking (1RSB) form 
taking values $q_1$ and $\mu_1$ on $n/m_1$ diagonal blocks 
(labelled by $B_k$, $k=1,2,\cdots, n/m_1$ of size $m_1$ and $q_0$ and
$\mu_0$ outside the blocks. 
We then have the terms in Eq.~(\ref{G}) as
\begin{align}
 &\sum_{a\neq b}q^2_{ab}=n[(m_1-1)q^2_1 +(n-m_1)q^2_0], \label{qp} \\
 &\sum_{a\neq b}\mu_{ab}q_{ab}= n[(m_1-1)\mu_1 q_1
 +(n-m_1)\mu_0 q_0].\label{lamp}
\end{align}
We will focus on the 1RSB solutions with $q_0=\mu_0=0$.
By writing
\begin{align}
&\frac 1 {2M}\sum^M_{i<j} \sum_{a\neq b} \mu_{ab}
 S^a_i S^a_j  S^b_i S^b_j  \nonumber \\
&= \frac {\mu_1} {2M}\sum_{k=1}^{n/m_1}  \sum^M_{i<j} \left\{ \left[ \sum_{a \in B_k} S^a_i S^a_j \right]^2 - m_1\right\} 
 \label{lss00}
\end{align}
in Eq.~(\ref{L}) and by using the Hubbard-Stratonovich transformation, we have
\begin{align}
&\underset{\{S_i^a\}}{\mathrm{Tr}}\, \exp\left[ \frac 1 {2M}\sum^M_{i<j} \sum_{a\neq b} \mu_{ab}
 S^a_i S^a_j  S^b_i S^b_j  \right]   \\
 =& \exp\left[-n \frac{\mu_1 K}{2M}\right]  \nonumber \\
&\times \Big[  \int D^K \bm{y}\; \Big\{\underset{\{S_i\}}{\mathrm{Tr}}\, \exp\Big[ \sqrt{\frac{\mu_1}M} \sum_{i<j}^M y_{(i,j)} 
S_i S_j \Big]\Big\}^{m_1}\Big]^{n/m_1}. \nonumber 
\end{align}
Therefore we have
\begin{align}
\lim_{n\to 0}\frac 1 n \ln L (\underline{\mu}) =& -\frac K {2M} \mu_1   +M\ln 2 \nonumber \\
&+\frac 1 {m_1}\ln \int D^K \bm{y}  \;  \zeta ^{m_1}(\bm{y},\mu_1)  , \label{lntr}
 \end{align}
 where $\zeta$ is defined in Eq.~(\ref{xidef}). 
Using Eqs.~(\ref{qp}), (\ref{lamp}) and (\ref{lntr}) in Eq.~({\ref{free1}), 
\begin{align}
 \frac{\beta F_{\rm 1RSB}}{N}=& -C \beta^2  -\frac M 4  \beta^2 (m_1-1)q^2_1 \nonumber \\ 
 &+  \frac M 2  (m_1-1) 
 \mu_1 q_1+\frac K {2M} \mu_1-M\ln 2\nonumber \\
 & 
-\frac 1 {m_1} \ln   \int D^K \bm{y}  \;  \zeta ^{m_1}(\bm{y},\mu_1) .
 \label{f_1rsb_1}
\end{align}

Varying the free energy with respect to $q_1$ and $\mu_1$, respectively, we have
\begin{equation}
 \mu_1=\beta^2 q_1.
 \label{lam_1rsb}
\end{equation} 
and 
\begin{align}
 q_1&=  \frac 1 {M^2} \frac{1}{ \int D^K \bm{y}\; \zeta^{m_1}(\bm{y},\mu_1)} \label{q1_1rsb}  \\
 &\times \int D^K \bm{y} \;  \zeta ^{m_1-2}
  \sum_{\alpha=1}^K\left\{ \frac 1 {2^M}\underset{\{S_i\}}{\mathrm{Tr}}\;  \Psi_\alpha \exp[ \sqrt{\frac{\mu_1}{M}} \bm{y}\cdot\bm{\Psi}] \right\}^2,
 \nonumber
\end{align}
Now varying the free energy with respect to $m_1$, we have
\begin{align}
&\frac M 4 \beta^2 q^2_1 +\frac 1 {m_1^2} \ln   \int D^K \bm{y}\; \zeta^{m_1} (\bm{y},\mu_1) \nonumber \\
&-\frac1 {m_1} \frac{\int D^K \bm{y}\; \zeta^{m_1}(\bm{y},\mu_1)\ln \zeta(\bm{y},\mu_1)}
{\int D^K \bm{y}\; \zeta^{m_1}(\bm{y},\mu_1)}=0. \label{m1_1rsb}
\end{align}
In summary, Eqs.~(\ref{lam_1rsb}), (\ref{q1_1rsb}), and (\ref{m1_1rsb}) are the saddle point equations
one has to solve for the 1RSB state.

Note that when $m_1=1$, we can explicitly evaluate 
\begin{align}
\int D^K \bm{y} \; \zeta(\bm{y},\mu_1)
= \exp\left[\frac K {2M} \mu_1\right].
\end{align}
From Eq.~(\ref{f_1rsb_1}), we see that when $m_1=1$, the 1RSB free energy is equal to the RS one:
\begin{equation}
 \frac{\beta F_{\rm 1RSB}}N \underset{m_1\to 1}{\rightarrow} - C\beta^2-M\ln2 
 = \frac{\beta F_{\rm RS}}N. 
 \end{equation}
To determine the transition temperature $T_c^{\rm 1RSB}$ to the 1RSB state, we set $m_1=1$ in 
Eqs.~(\ref{lam_1rsb}), (\ref{q1_1rsb}) and (\ref{m1_1rsb})
and solve for $\beta$.
For $m_1=1$, we can  combine 
these three equations into one equation, $f_M(\sigma)=0$ for the parameter
\begin{align}
\sigma\equiv\sqrt{\frac{\mu_1}M},
\end{align}
where
\begin{align}
f_M(\sigma)&\equiv e^{-K\sigma^2 /2} \int D^K \bm{y}\; \Big[ \zeta (\bm{y},\mu_1)\ln \zeta (\bm{y},\mu_1) 
\label{eq_m}\\
&-\frac {\sigma^2} {4} 
\frac{ \sum_{\alpha=1}^K\left\{ 2^{-M}\mathrm{Tr}\; \Psi_ \alpha \exp\left[ \sigma\bm{y}\cdot\bm{\Psi}\right] \right\}^2}{\zeta(\bm{y},\mu_1)}\Big]
-\frac K {2} \sigma^2.
\nonumber 
\end{align}
Note that $\zeta(\bm{y},\mu_1)$ is a function of $\sigma$. If there exists a nonzero solution $\sigma$ to $f_M(\sigma)=0$, 
one can obtain
nonzero $q_1$ from Eq.~(\ref{q1_1rsb}) and the transition temperature $T_c^{\rm 1RSB}$ from Eq.~(\ref{lam_1rsb}).

 \begin{figure}
  \includegraphics[width=0.85\columnwidth]{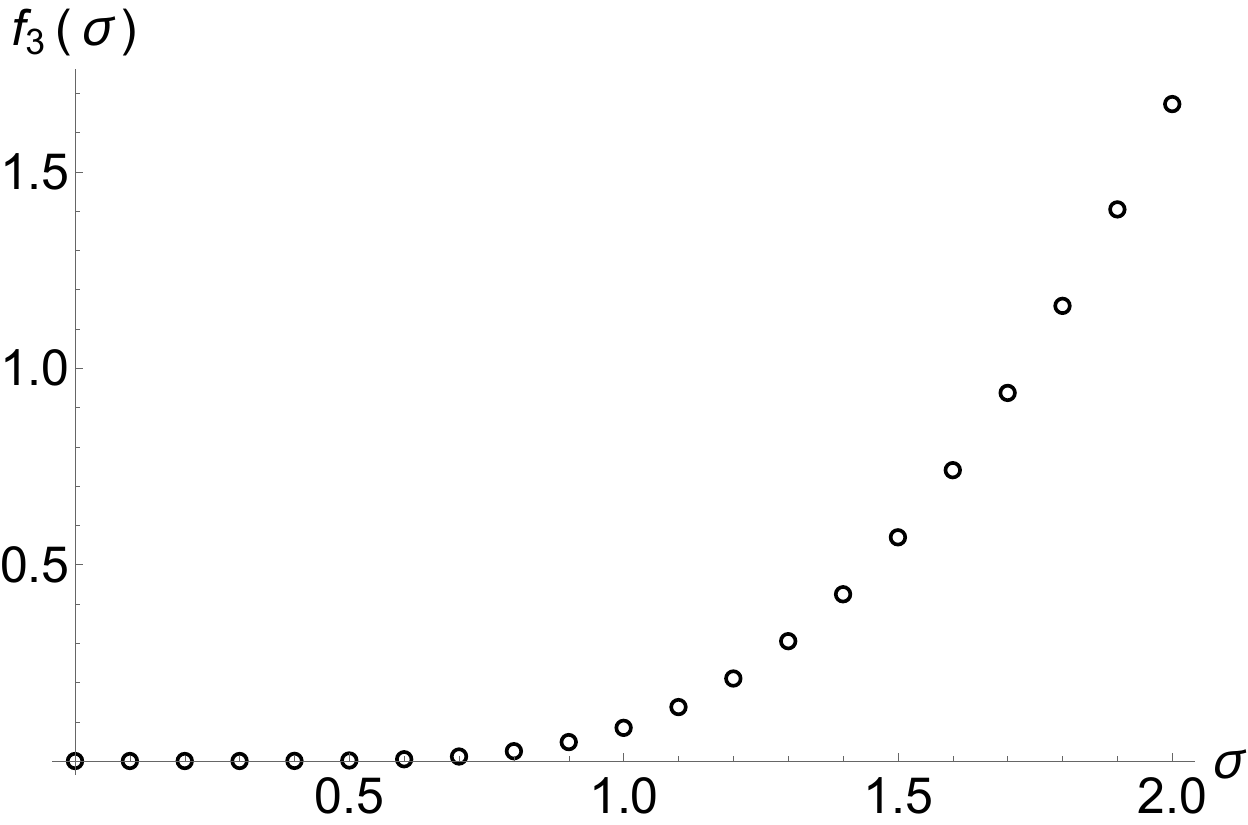}
  \caption{$f_M(\sigma)$ defined in Eq.~(\ref{eq_m}) for $M=3$. A nonzero solution $\sigma$ of
  $f_M(\sigma)=0$ would signal a discontinuous transition into the 1RSB state. }
  \label{fig:m3}
\end{figure}

\begin{figure}
  \includegraphics[width=0.9\columnwidth]{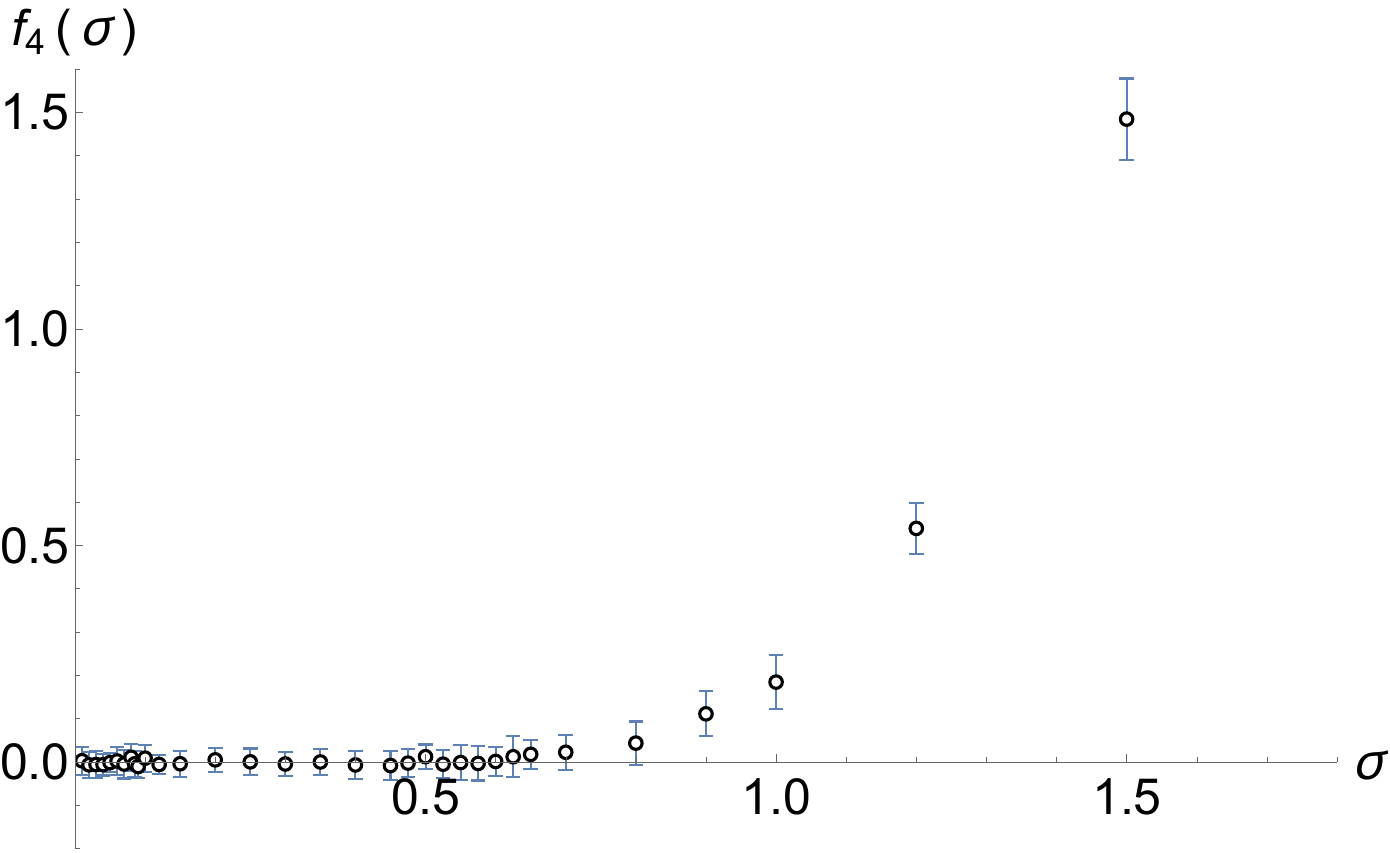}
  \caption{Same as Fig.~\ref{fig:m3} with $M=4$.}
  \label{fig:m4}
\end{figure}

We solve this equation by numerically evaluating multi-dimensional integrals in Eq.~(\ref{eq_m}). 
In Figs.~\ref{fig:m3} and \ref{fig:m4}, $f_M$ is plotted as a function of $\sigma$
for $M=3$ and $M=4$.
As we can see from the figures, $f_M(\sigma)$ starts off very flat and increases monotonically for large values of
$\sigma$. For $M=3$, Fig.~\ref{fig:m3} clearly shows a monotonic increase as a function of $\sigma$, thus we can conclude that
the only solution to $f_3(\sigma)=0$ is $\sigma=0$. From Eq.~(\ref{q1_1rsb}), we then have $q_1=0$ thus no
discontinuous transition in this case. 
For $M=4$, we have to evaluate six-dimensional ($K=6$) integrals in Eq.~(\ref{eq_m}).
For that, we use Monte Carlo methods, and
the results are shown in Fig.~\ref{fig:m4}.
The error bars come from sampling random points in the integrands within the Monte Carlo evaluation of the integrals.
We have averaged over 30 trials for each data point.
Since $f_4(\sigma)$ stays very flat for small $\sigma$ before increasing to large positive values, 
it is quite difficult to determine, if any, nonzero solution $\sigma$ from this plot alone.  

To understand the situation more clearly, we study the behavior of $f_M(\sigma)$ for small $\sigma$. 
We can show (see Appendix \ref{app:smalls} for details) that for small $\sigma$, the leading order in
the small-$\sigma$ expansion of $f_M(\sigma)$ is $O(\sigma^6)$. In fact, if we write
$f_M(\sigma)=\sum_{i=0}^\infty c_{i}(M) \sigma^i$, we find that $c_i=0$ for $i$ odd,
$c_0=c_2=c_4=0$ and 
\begin{align}
c_6 (M)= -\frac M{24} (M-1)(M-3)
\label{c6}
\end{align}
for $M=3,4,5,\cdots$. Therefore, for $M=3$, the leading order is actually $O(\sigma^8)$. 
The next-order coefficient is given by
\begin{align}
c_8(M)=-\frac{M}{48} (M-1)(3M^2-27M+47),
\label{c8}
\end{align}
for $M\ge 3$.
Some steps needed to obtain these are given in Appendix \ref{app:smalls}.
We note that $c_8 (M=3) >0$. This is consistent with the monotonic increase of $f_3(\sigma)$ shown in 
Fig.~\ref{fig:m3}. For $M>3$, $c_6$ becomes negative. Combining this fact with the 
monotonic increase for large $\sigma$, we can conclude that there exists a nonzero solution to $f_M(\sigma)=0$
and that a discontinuous transition for $M>3$ is expected. 
From Eq.~(\ref{c8}), we find that $c_8(M)>0$ for $M\lesssim 6.64$, therefore for these values of $M$, 
we can estimate the solution as $\sigma\simeq \sqrt{-c_6(M)/c_8(M)}$.
This program, however, fails when $c_8(M)<0$ for $M\gtrsim 6.64$. ($c_6<0$ for $M>3$.) 

We need to go to higher order to study the 1RSB transition beyond this value of $M$. We find, however, that the method in Appendix \ref{app:smalls}
becomes too cumbersome to get $c_{10}$. The Landau expansion of the free energy given in Eq.(\ref{quinticLG}) provides a more
useful tool. Since $\sigma^2\sim\mu_1\sim q_1$, $O(\sigma^6)$ and $O(\sigma^8)$ correspond to the cubic and quartic orders in $q_{ab}$,
respectively, and
we need quintic order terms in $q_{ab}$ to evaluate $c_{10}$. 
In Appendix \ref{app:1rsb}, we apply the 1RSB form directly to $q_{ab}$ 
in Eq.~(\ref{quinticLG}).
When $m_1=1$, the saddle point equations can be combined into a form 
\begin{align}
-\frac{1}{2}(w_2-w_1)q^3_1- (y_1-y_3+y_5) q_1^4 -\frac{3}{2} z_1^{\rm eff} q^5_1=0 ,
\label{T1rsbeq}
\end{align}
where 
\begin{align}
z_1^{\rm eff}\equiv z_1-z_3-z_4+z_5+z_8-z_9.
\end{align}
Recalling that $q_1=\mu_1/\beta^2 =M\sigma^2/\beta^2$ and using the values of $w_i$ and $y_i$ given in 
Appendix \ref{app:exp}, we can identify the first two terms in Eq.~(\ref{T1rsbeq}) as the small-$\sigma$ expansion of $f_M(\sigma)$, since we
can rewrite
\begin{align}
c_6(M)=-\frac{M^3}{2\beta^6}(w_2-w_1), 
\end{align}
and
\begin{align}
c_8(M)=  -\frac{M^4}{\beta^8} (y_1-y_3+y_5).
\end{align}
It follows that the last term in Eq.~(\ref{T1rsbeq}) gives
\begin{align}
c_{10}(M)=-\frac{3M^5}{2\beta^{10}}z_1^{\rm eff}.
\end{align}
The explicit expression as a function of $M$ is given in Eqs.~(\ref{app:z1eff}) and (\ref{app:c10}) in Appendix \ref{app:1rsb}.

In Fig.~\ref{fig:zeff}, $(y_1-y_3+y_5)/\beta^{8}$ and $z_1^{\rm eff}/\beta^{10}$ are displayed as functions of $M$.
We note that $y_1-y_3+y_5$ is negative (and $c_8$ is positive) for $2.35 \lesssim M \lesssim 6.64$.   
Therefore, as we mentioned above, we can find the 1RSB solution for $m_1=1$ for $3<M\lesssim 6.64$ within the quartic theory. 
The result for the 1RSB transition temperature obtained in this way is shown as a solid red line in Fig.~\ref{fig:temp} (a). 
We note, however, that the result becomes unreliable as we approach the boundary value $M\simeq 6.64$ as it shows 
a fictitious diverging behavior. We now study how the quintic theory may improve this result.
The quintic contribution can be summarized by $z_1^{\rm eff}$, which is negative for $4.37\lesssim M\lesssim 12.46$
(and for the narrow region $2\le M\lesssim 2.12$). Since $c_{10}$ is positive in that interval, we have a chance to extend 
the result of the quartic theory to larger values of $M$. As one can see in Fig.~\ref{fig:temp} (a), the 1RSB transition line 
calculated within the quintic theory indeed extends to large values of $M$. But, since Eq.~(\ref{T1rsbeq})
for $q_1\neq 0$ becomes a quadratic equation for $q_1$, there are intervals of $M$ where
no real solution exists. We find that for $3.27\lesssim M\lesssim 3.98$ and for $M\gtrsim 14.41$,
solutions to this equation become complex and no 1RSB solution can be obtained. This can be seen in Fig.~\ref{fig:temp} (b),
where one can see a segment of the 1RSB transition line is missing. Also as in the quartic theory, the transition
line displays an apparent divergent behavior as we approach the boundary value $M\simeq 14.41$.
Therefore, we can conclude that it is possible to obtain the 1RSB transition line using truncated models, 
but the truncation of the free energy to
a specific order produces some unphysical features. 
Comparing the results of 
the quartic and quintic theories in Fig.~\ref{fig:temp} (a), 
we expect that a systematic improvement may occur if we go to even higher orders. 
We also note that the 1RSB transition temperatures obtained in this way always stay above $T_*$.

The 1RSB transition line discussed above is obtained by setting $m_1=1$ where the 1RSB free energy coincides
with that of the high-temperature RS phase (with $q=0$). 
Using the results in Appendix \ref{app:1rsb}, we can obtain 1RSB solutions 
for general values of $0\le m_1 \le 1$ for the truncated model. Rather unexpectedly, we find that for given $M$, the 1RSB solution ceases to 
exist below a certain finite temperature for which $m_1=0$. We note that if $m_1=0$, the 1RSB free energy 
becomes that of the RS phase with nonzero $q$ (see Eq.~(\ref{LGfree1rsb})). Therefore, below that temperature, we
only have the RS solution with nonzero $q$. This is illustrated in Fig.~\ref{fig:free},
where we plot the free energies of both 1RSB and RS solutions calculated within a truncated model.
One can clearly see that the 1RSB solution exists only in a finite temperature interval.
Within that interval, the system is in the 1RSB phase which has a higher free energy than the RS one with nonzero $q$.
However, below that interval, there is no 1RSB solution, so the system returns to the RS phase.
We believe that this rather unusual behavior is caused by the
truncation of the model in an arbitrary order. In the large-$M$ limit considered in Sec.~\ref{sec:largem}, 
where one can find the 1RSB solutions
without truncation, we find that the 1RSB solution continues down to zero temperature and has a higher free energy 
than the RS one. 

\begin{figure}
  \includegraphics[width=0.95\columnwidth]{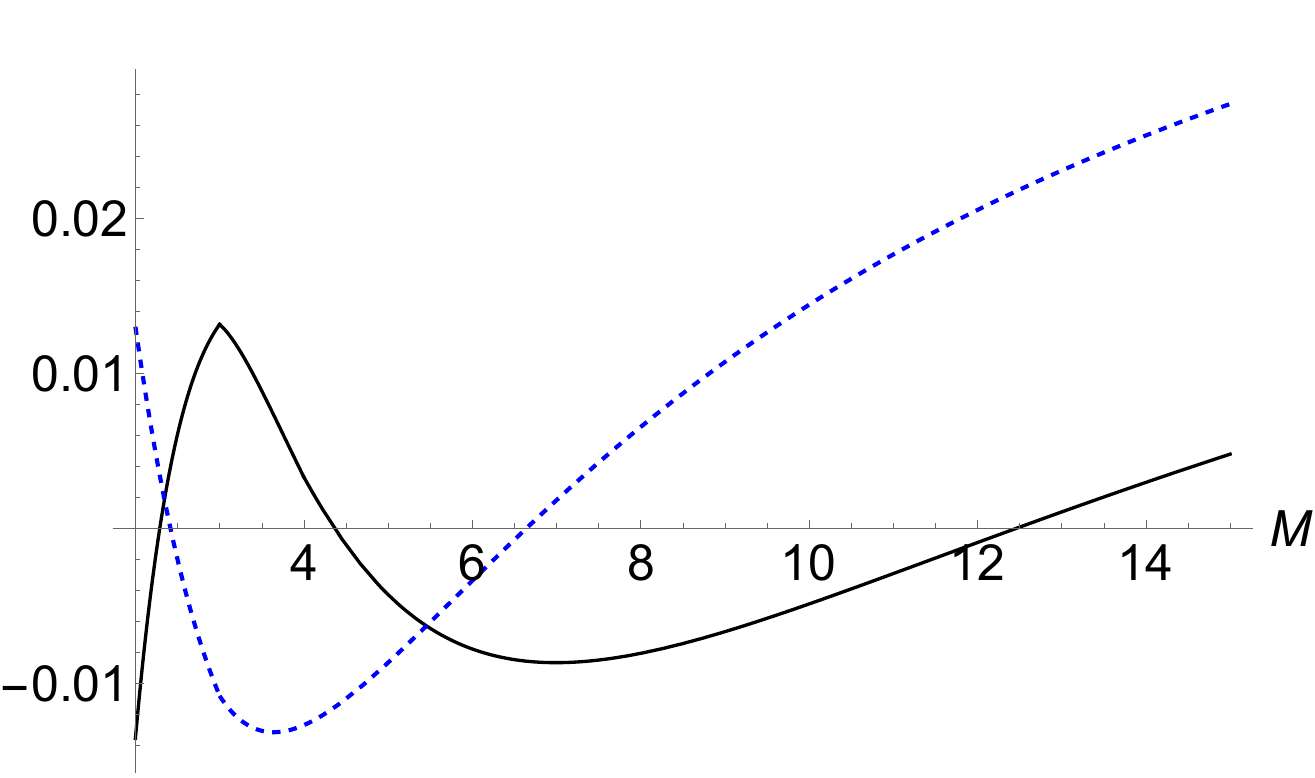}
  \caption{$(y_1-y_3+y_5)/\beta^{8}$ (dashed line) and $z_1^{\rm eff}/\beta^{10}$ (solid line) as functions of $M$.
  In the large-$M$ limit, they approach 1/16 and 1/20, respectively.}
  \label{fig:zeff}
\end{figure}

\begin{figure}
  \includegraphics[width=0.95\columnwidth]{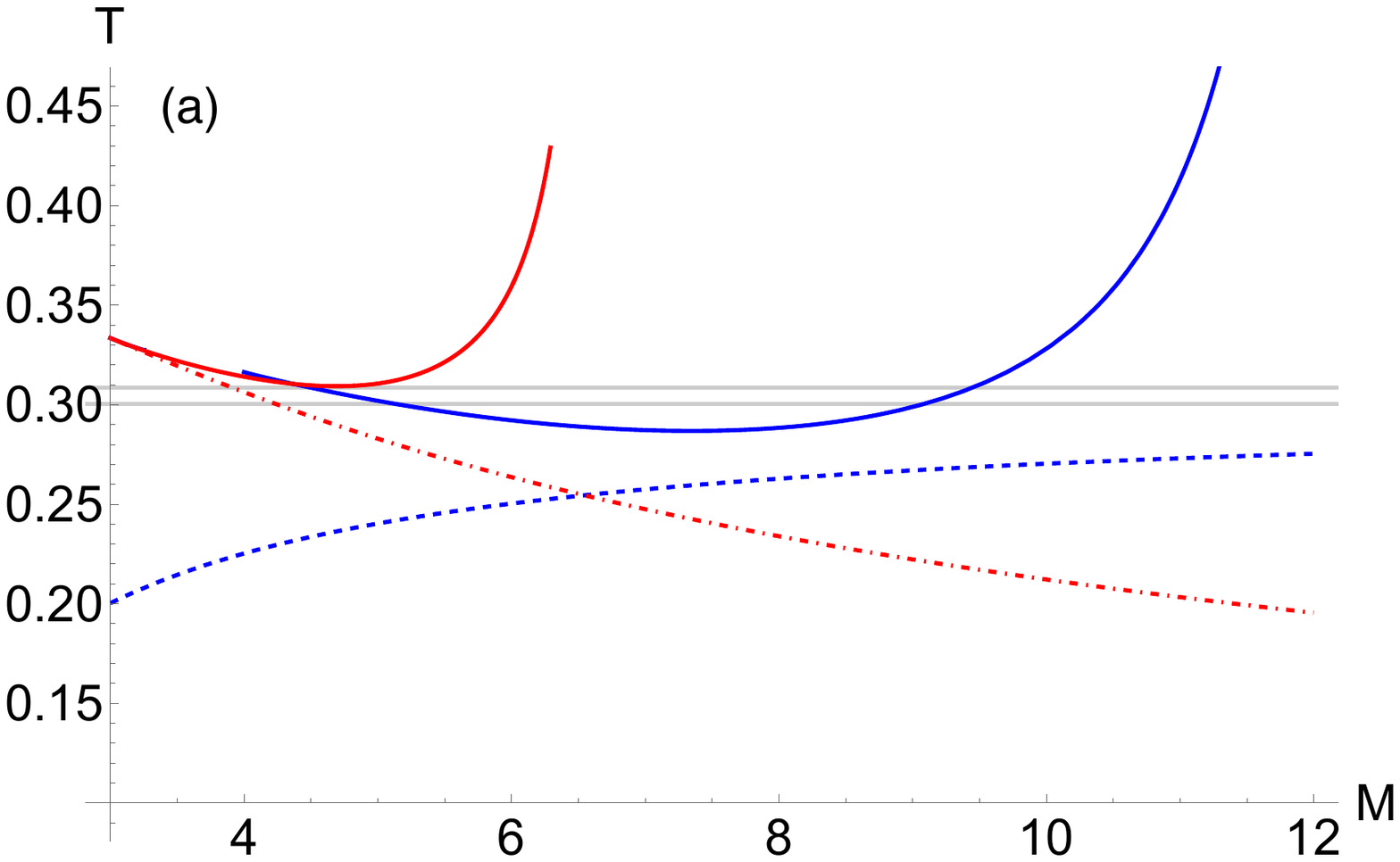}
   \includegraphics[width=0.95\columnwidth]{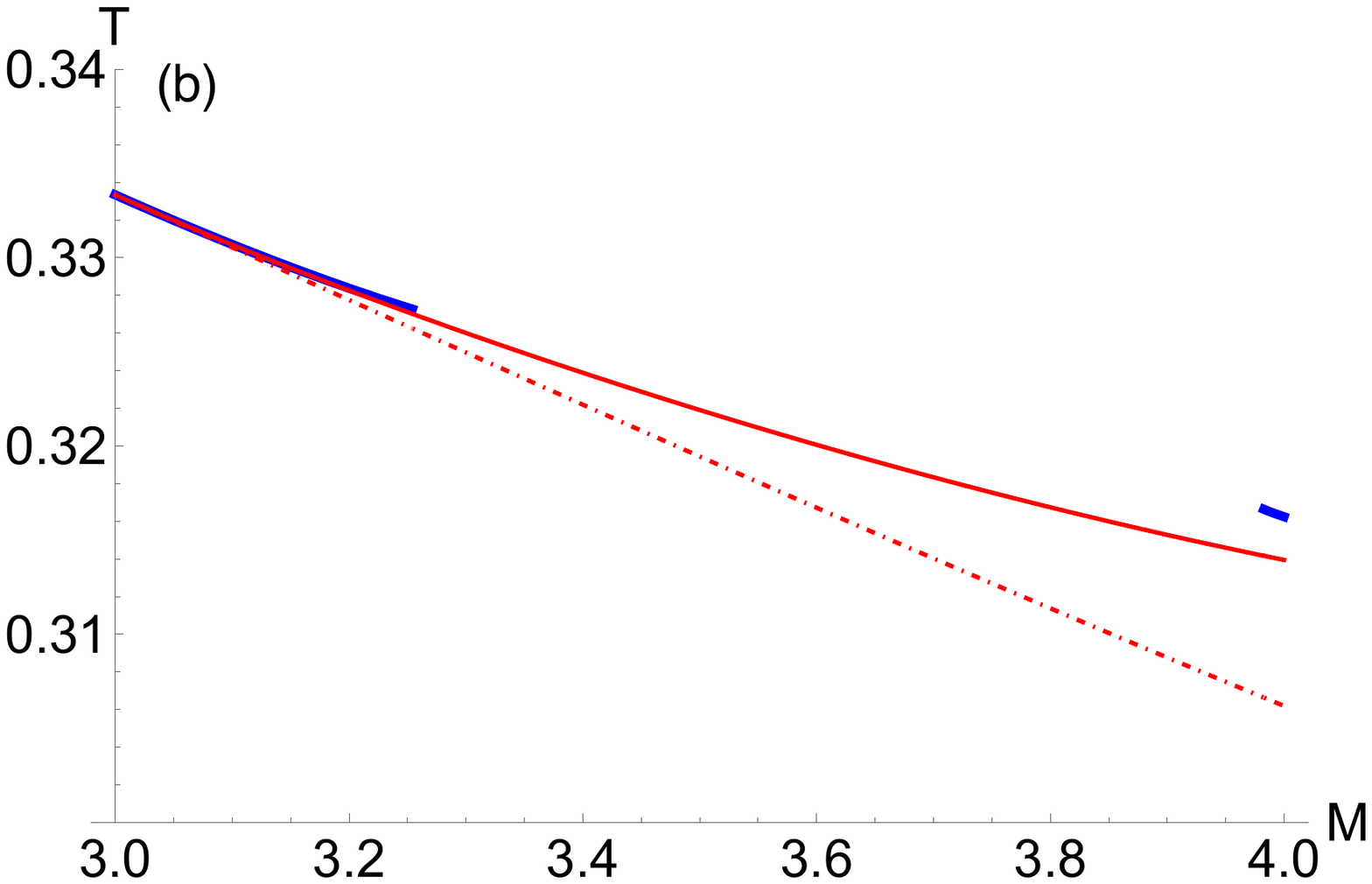}
  \caption{(a) Red and blue solid lines are the 1RSB transition temperatures $T_c^{\rm 1RSB}$ 
  as functions of $M$ for the $p=4$ balanced $M$-$p$ model
  expanded up to quartic (red) and to quintic (blue) orders in the order parameter.
  Dashed and dot-dashed lines are $T_*$ (Eq.~(\ref{tstarm})) and $T_c$ (Eq.~(\ref{Tc})), respectively.   
  Two closely-spaced horizontal lines are the large-$M$ limits of $T_*$ (lower one, Eq.~(\ref{tsinfty})) and 
  $T_c^{\rm 1RSB}$ (upper one, Eq.~(\ref{t1rsbinfty})).
  (b) Close-up of the same plot for $3\le M\le 4$. There is a gap in the solid blue line
  in the interval $3.27\lesssim M\lesssim 3.98$, where no 1RSB solution exists at $m_1=1$
  for the quintic theory. The red line corresponds to the quartic theory, which has no gap. 
  The dot-dashed line is $T_c$.}
  \label{fig:temp}
\end{figure}

\begin{figure}
  \includegraphics[width=0.95\columnwidth]{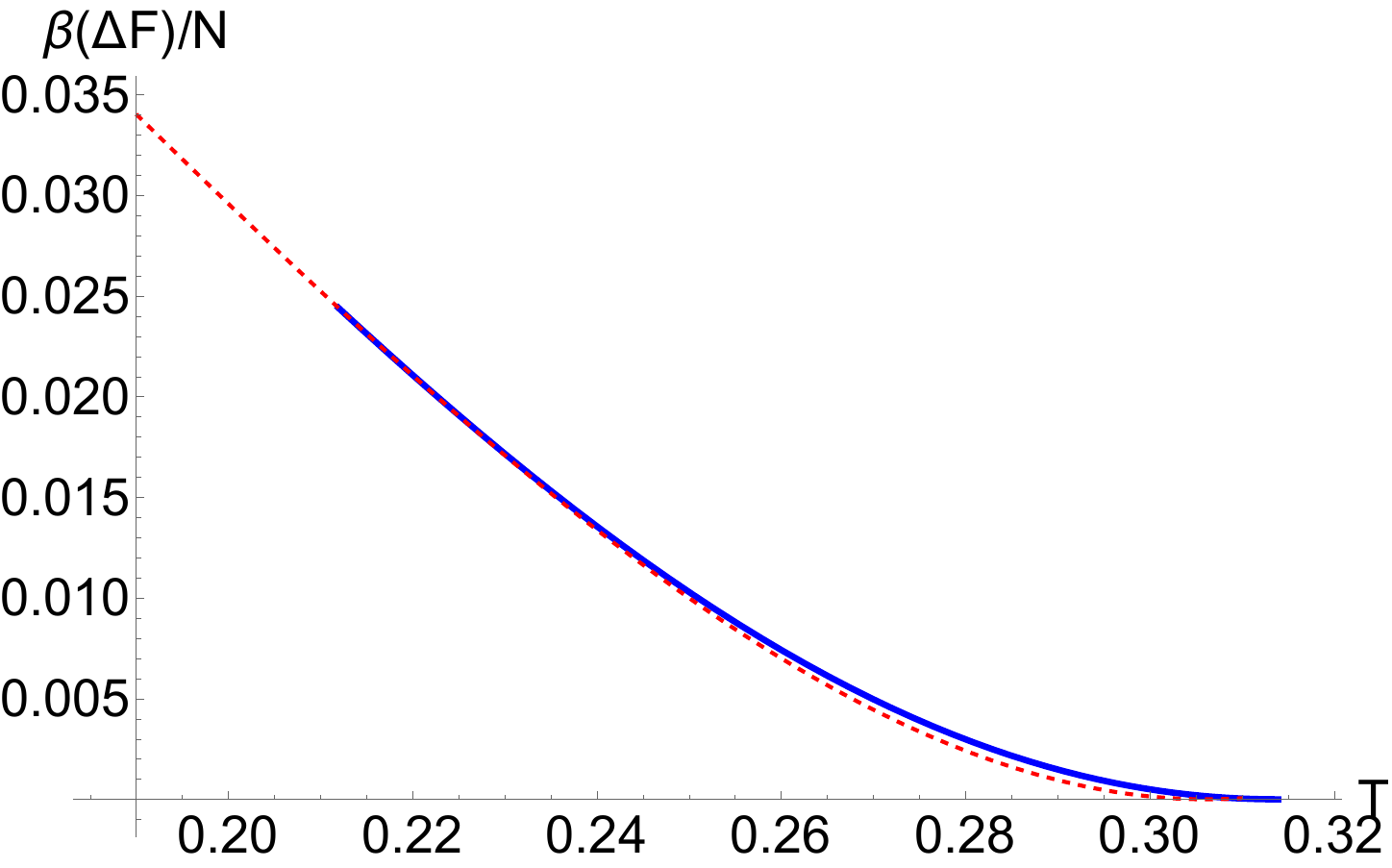}
  \caption{Dimensionless free energies per spin of the 1RSB solution (solid line) and the RS solution with $q \neq 0$ (dashed line)
  as functions of temperature calculated for  
  the quartic $M=4$ model. 
  For each case, the free energy difference ($\Delta F$) from that of the high-temperature RS solution ($q=0$, Eq.~(\ref{free_highT}))
  is plotted. The 1RSB solution exists only in the temperature interval $0.212 \le T \le 0.311$. }
  \label{fig:free}
\end{figure}

\subsection{The Large-$M$ Limit}
\label{sec:largem}

In this subsection, we consider the situation where we take the limit $M\to\infty$ from the start.
In the large-$M$ limit, Eq.~(\ref{L}) can be rewritten as
\begin{align}
\label{L_minfty}
L(\underline{\mu})&=\underset{\{S_i^a\}}{\mathrm{Tr}} \exp\left[   \frac 1 {4M} \sum_{a\neq b} \mu_{ab}\left\{
\left( \sum^M_i
 S^a_i S^b_i  \right)^2-M \right\} \right]\nonumber \\
 &\simeq \underset{\{S_i^a\}}{\mathrm{Tr}} \exp\left[   \frac M {4} \sum_{a\neq b} \mu_{ab} 
\left( \frac 1 M \sum^M_i
 S^a_i S^b_i  \right)^2 \right],
\end{align}
where we have neglected the subleading terms in the large-$M$ limit.
We now introduce the delta function
$\delta (MQ_{ab}-\sum_i^M S_i^aS_i^b)$
using the integral representation with the variable $\lambda_{ab}$. Then we have from Eq.~(\ref{zn})
\begin{align}
\overline{Z^n}=&e^{nN\beta^2 C}\int\prod_{a<b} dq_{ab}  d\mu_{ab} dQ_{ab} d\lambda_{ab} 
 \nonumber  \\
 \times&\exp\Big[-NM\Big\{ 
 - \frac 1 4 \beta^2\sum_{a\neq b}q^2_{ab}
+\frac 1 2\sum_{a\neq b}\mu_{ab}q_{ab}\nonumber \\
& - \frac 1 4 \sum_{a\neq b}\mu_{ab}Q^2_{ab}+\frac 1 2 \sum_{a\neq b} \lambda_{ab} Q_{ab}
-\ln \tilde{L}(\underline{\lambda})
\Big\}\Big] \label{zn_minfty}
\end{align}
where
\begin{align}
\widetilde{L}(\underline{\lambda})=\underset{\{S^a\}}{\mathrm{Tr}} \exp\left[   \frac 1 {2} \sum_{a\neq b} \lambda_{ab} 
 S^a S^b\right].
\end{align} 
In the large-$M$ limit, the integral is dominated by the saddle points. In particular, the saddle point equations obtained by varying $q_{ab}$ 
and $\mu_{ab}$ are, respectively,
\begin{align}
\mu_{ab}=\beta^2 q_{ab}
\end{align}
and
\begin{align}
q_{ab}=\frac 1 2 Q^2_{ab}.
\end{align}
Inserting this into the above equation, we can rewrite Eq.~(\ref{zn_minfty}) as
\begin{align}
\overline{Z^n}=e^{nN(\beta J)^2 C}\int\prod_{a<b} dQ_{ab} d\lambda_{ab} 
\;\exp[-NM\widetilde{G}(\underline{Q},\underline{\lambda})],
\label{zn_minfty1}
\end{align}
with
\begin{equation}
\widetilde{G}(\underline{Q},\underline{\lambda})=
- \frac 1 {16} ( \beta J)^2\sum_{a\neq b}Q^4_{ab}+\frac 1 2 \sum_{a\neq b} \lambda_{ab} Q_{ab}-\ln \tilde{L}(\underline{\lambda}).
\label{G_minfty}
\end{equation}
The free energy in the large-$M$ limit is then given by
\begin{align}
\frac{\beta F}{NM}=-(\beta J)^2 C_{\infty} +\lim_{n\to 0}\frac 1 n \widetilde{G}(\underline{Q},\underline{\lambda}),
\label{free_minfty}
\end{align}
where
\begin{align}
C_{\infty}=\lim_{M\to\infty}\frac{C}{M}=\frac 1 {16}
\label{c_infty}
\end{align}
Note that we have restored $J^2$ which sets the variance in Eq.~(\ref{variance}) explicitly.
This free energy is exactly the same as the one for the fully connected $p$ spin glass model with $p=4$, which is given by the Hamiltonian
\begin{align}
H=-\sum_{1\le x_1<\cdots<x_p\le N}J_{x_1,x_2,\cdots,x_N}S(x_1)S(x_2)\cdots S(x_p),
\end{align}
for the Ising spin $S(x)$ at site $x$. The bonds $J_{x_1,x_2,\cdots,x_N}$ are independent random variables 
satisfying the Gaussian distribution with zero mean and variance
\begin{align}
\frac{p ! \tilde{J}^2}{2N^{p-1}} .
\end{align}
The free energy for this model is given exactly the same as Eq.~(\ref{free_minfty}) with $\tilde{J}^2=J^2/4$. (The formula
for this correspondence for general $p$ is $\tilde{J}^2=4C_{\infty}J^2$.)

We can readily use the known results for this model. The replica symmetric phase with $\lambda=Q=0$ has
the free energy per site as
\begin{equation}
  \frac{\beta F_{\rm RS}}N= -\frac{(\beta \tilde{J})^2}{4}-\ln 2.
 \label{f_rs}
\end{equation}
The entropy per site is then given by
\begin{equation}
 \frac{S_{\rm RS}}N=\ln 2- \frac {(\beta \tilde{J})^2}{4},
\end{equation}
which becomes negative for temperature $T/\tilde{J}<T^\infty_*/\tilde{J}\equiv 1/(2\sqrt{\ln 2})$. Therefore in the original unit 
\begin{align}
T^\infty_*/J=\frac{1}{4\sqrt{\ln 2}}\simeq 0.30028 .
\label{tsinfty}
\end{align}
This is the same value as that obtained in the $M\to\infty$ limit of Eq.~(\ref{tstarm}).

If we use the 1RSB form for $Q_{ab}$ and $\lambda_{ab}$ in Eq.~(\ref{free_minfty}), the free energy becomes
\begin{align}
 \frac{\beta F^\infty_{\rm 1RSB}}{N}=& -\frac {(\beta \tilde{J})^2}{4} [1+(m_1-1)Q_1^p] +\frac 1 2 (m_1-1)\lambda_1 Q_1
 \nonumber \\
 +&
 \frac{\lambda_1}{2}   -\ln 2 -\frac 1 { m_1}
\ln \int Dy\; \cosh^{m_1}(\sqrt{\lambda_1} y) .
\label{f_1rsb_2_inf}
\end{align}
The saddle point equations are
as follows:
\begin{equation}
 \lambda_1=\frac {(\beta \tilde{J})^2}2 p Q_1^{p-1},
 \label{lambda1}
\end{equation}
and 
\begin{equation}
 Q_1= \frac{\int Dy\; \cosh^{m_1} (\sqrt{\lambda_1}y)\tanh^2 (\sqrt{\lambda_1}y)}
 {\int Dy\; \cosh^{m_1}(\sqrt{\lambda_1}y)}.
 \label{q1_1rsb_inf}
\end{equation}
There is another saddle point equation which is obtained by varying the free 
energy with respect to $m_1$:
\begin{align}
&\frac{(\beta \tilde{J})^2}4 Q_1^p (p-1) + \frac 1 {m_1^2} \ln \int Dy\;
 \cosh^{m_1} (\sqrt{\lambda_1}y)  \nonumber \\
&~~ -\frac 1 {m_1} \frac{\int Dy\; \cosh^{m_1} (\sqrt{\lambda_1}y)
\ln(\cosh(\sqrt{\lambda_1}y))}{\int Dy\; \cosh^{m_1} (\sqrt{\lambda_1}y)}=0.
\label{m1_1rsb_inf}
 \end{align}
 Again, when $m_1=1$, $F_{\rm 1RSB}$ becomes equal to $F_{\rm RS}$.
We determine the temperature $T^\infty_{\rm 1RSB}$ by setting $m_1=1$.
Using $\int Dy\; \cosh(\sqrt{\lambda_1}y)=e^{\lambda_1/2}$, we can combine Eqs.~(\ref{q1_1rsb_inf}),
(\ref{m1_1rsb_inf}) and (\ref{lambda1}) to get
\begin{align}
& e^{-\lambda_1/2}\int Dy\; \cosh(\sqrt{\lambda_1}y)\Big[ 
 \ln\cosh(\sqrt{\lambda_1}y) \nonumber \\
 &~~~~~~~~~~ -\frac{p-1}{2p}\lambda_1 \tanh^2(\sqrt{\lambda_1}y)\Big]
 -\frac{\lambda_1}{2}=0.
\end{align}
If we define
\begin{align}
\nu \equiv \sqrt{\lambda_1},
\end{align}
then the above equation can be rewritten as $f_\infty(\nu)=0$ where
\begin{align}
f_{\infty}(\nu)\equiv & ~ e^{-\nu^2/2}\int Dy\; \Big[ \cosh(\nu y)
 \ln\cosh(\nu y) \nonumber \\
 &~~~~~~~~~~~~~ -\frac{p-1}{2p}\nu^2 \frac{\sinh^2(\nu y)}{\cosh(\nu y)}\Big]
 -\frac{\nu^2}{2}. 
 \label{f_infty}
\end{align}
This is to be compared with the corresponding Eq.~(\ref{eq_m}) for finite $M$. 
In Fig.~\ref{finfty}, $f_\infty(\nu)$ is plotted for $p=4$. From the nonzero solution and from the corresponding $Q_1$ in
Eq.~(\ref{q1_1rsb_inf}) and the relation Eq.~(\ref{lambda1}), we obtain $T^\infty_{\rm 1RSB}/\tilde{J}\simeq 0.61688$ or
in the original unit
\begin{align}
T^\infty_{\rm 1RSB}/J\simeq 0.30844 >T^*_\infty.
\label{t1rsbinfty}
\end{align}
For $f(\nu)$, the small-$\nu$ expansion yields
\begin{align}
f_\infty(\nu)=& \left( \frac{2-p}{4p}\right)\nu^4 +\left( \frac{2p-3}{6p}\right) \nu^6 \nonumber \\
&+\left( \frac{5(4-3p)}{24p}\right)\nu^8 +O(\nu^{10}).
\end{align}
We can see that for $p>2$, $f_\infty(\nu)$ has a negative slope near the origin. For $p=2$, the leading order term is $\nu^6$ with a positive coefficient.

\begin{figure}
 \includegraphics[width=0.9\columnwidth]{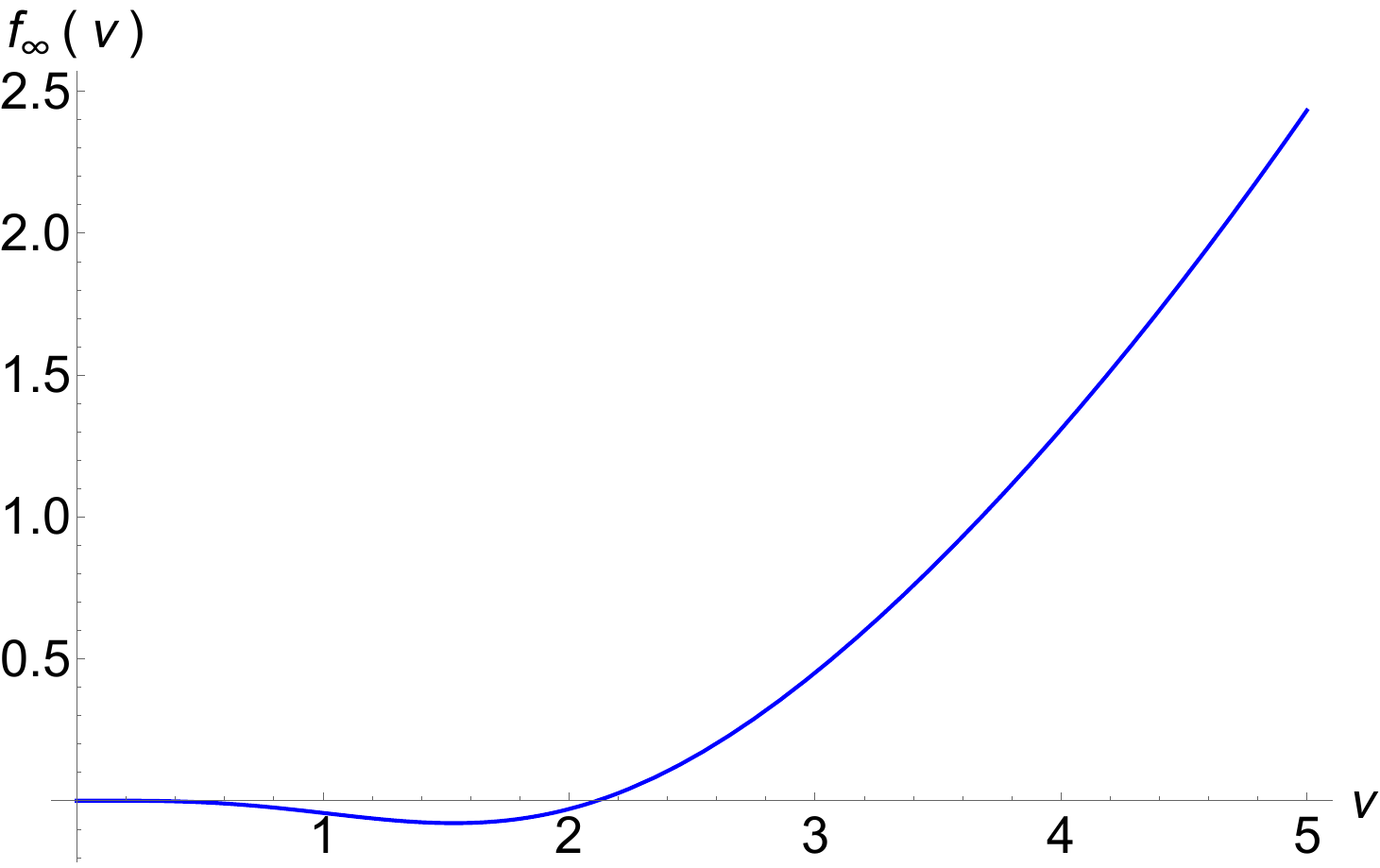}
 \caption{$f_\infty(\nu)$ defined in Eq.~(\ref{f_infty}) for $p=4$. There is a nonzero solution $\nu\simeq 2.1163$
 to the equation $f_\infty(\nu)=0$. }
 \label{finfty}
 \end{figure}

\subsection{The FRSB Solution}
\label{sec:frsb}
Here we consider the FRSB solutions. We first write the free energy in terms of the Parisi function 
$q(x)$ for $0\le x \le 1$. It is given by 
\begin{align}
&\frac{\beta F_{\rm FRSB}}{N}=   -C\beta^2  -M\ln 2  -\tau \langle q^2 \rangle  \nonumber \\
& - w_1 \int_0^1 dx\; \left\{ x q^3(x) +3 q(x)\int_0^x dy\; q^2(y) \right\} 
+ w_2 \langle q^3 \rangle  \nonumber \\ 
&+  y_1 \langle q^4 \rangle +
y_2 \Big\{ \langle q^4 \rangle -2 \langle q^2 \rangle^2 \nonumber \\
&~~~~~~~~~~~~~ -\int_0^1 dx\int_0^x dy\; (q^2(x)-q^2(y))^2 \Big\} \nonumber \\
&-y_3 \Big\{ 2\langle q \rangle\langle q^3 \rangle + \int_0^1 dx\; q^2(x) \int_0^x dy\; (q(x)-q(y))^2 \Big\} \nonumber \\
&-y_5 \Big\{ \langle q^2 \rangle^2 -4\langle q \rangle^2\langle q^2 \rangle \nonumber \\
&~~~~~  -4\langle q \rangle 
\int_0^1dx\; q(x)\int_0^x dy\; (q(x)-q(y))^2 \nonumber \\
&~~~~~ -\int_0^1 dx \int_0^x dy \int_0^x dz\; (q(x)-q(y))^2(q(x)-q(z))^2 \Big\}  \nonumber \\
&+z_1 \langle q^5 \rangle 
, \label{FRSB}
\end{align}
where
\begin{align}
\langle q^k \rangle =\int_0^1 q^k(x) dx.
\end{align}
and we have only kept the first quintic term. The FRSB expressions for the rest of the quintic terms are 
given in Appendix \ref{app:frsbquintic}.

Because the equations for the stationarity equations  of the FRSB functional equations are so cumbersome we have relegated them to the Appendices \ref{app:frsbeq} and \ref{app:frsbquintic}. We  can only make progress in solving these equations at the quintic level by making simplifications. The full set of quintic terms is given in Appendix \ref{app:frsbquintic} but in Eq.~(\ref{FRSB}) we have reduced them from 9  terms to just one. A similar device was used by Parisi \cite{Parisi_1980} at quartic level when he retained only the $y_1$ term. Subsequent studies have shown that the physics was hardly changed by such an approximation, but numerical values do get modified. We choose the numerical value of that $z_1$ to equal $z_1^{\rm{eff}}$ in Eq.~(\ref{app:z1eff}). A second simplification was to set $y_5$ =0. When this is done the differential equation of Eq.~(\ref{qprime0}) can be solved analytically. With $y_5$ set to be zero we do not think that does much harm to the physics of the problem. For example, the Goldbart-Elderfield singularity \cite{goldbart:85} still arises. But without the approximations of retaining only the $z_1$ term and setting $y_5$ to zero, the numerical work required for a solution would have been much harder.

Fortunately at quartic level, that is, if we set $z_1=0$, one can solve the differential equation for $q(x)$, Eq.~(\ref{qprime0}), analytically.  There is no need to set $y_5$ to zero when just working at quartic level. Because it is a first order differential equation, its solution depends on one adjustable constant $x_0$. The result is 
\begin{align}
q(x)=\frac{w_1 y_3 - 2 w_2 y_5 - \frac{2 (y_3 - 2 x y_5) (y_3^2 - 4 y_1 y_5) x_0}{\sqrt{
  y_1 - xy_3 + x^2 y_5}}}{2 (-y_3^2 + 4 y_1 y_5)}.
  \label{q(x)quartic}
\end{align}

\begin{figure}
 \includegraphics[width=0.9\columnwidth]{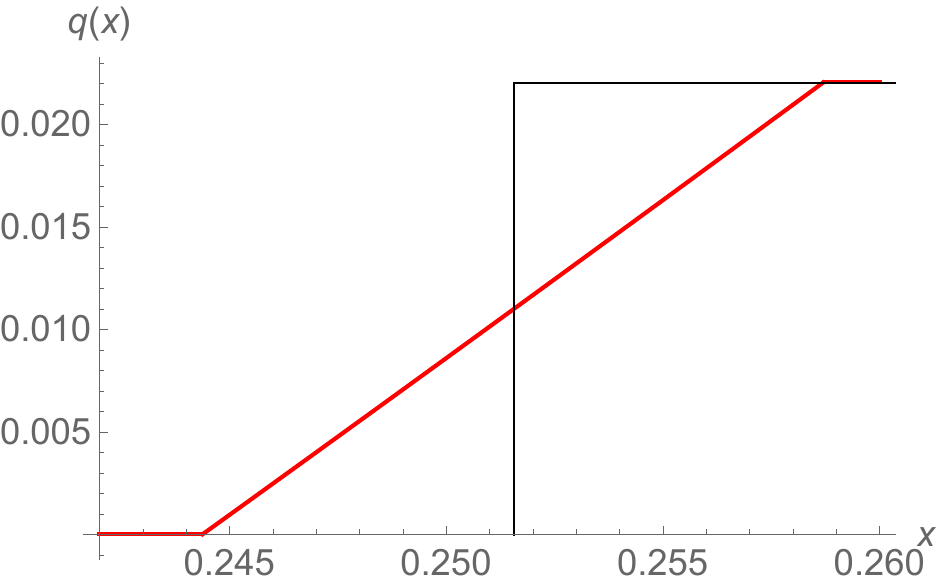}
 \caption{Plots of $q(x)$  for the FRSB solution (red)  and the 1RSB solution (black) at $M =2.25$  at $\tau=-0.001$. The FRSB state is the equilibrium state as it has the higher free energy. These plots are for the quartic theory. $x_1$ for the FRSB solution is where $q(x)$ goes to zero, $x_1\approx 0.24437$, while the upper breakpoint $x_2 \approx 0.25870$.}
 \label{q(x)M2.25}
 \end{figure}
 
 \begin{figure}
 \includegraphics[width=0.9\columnwidth]{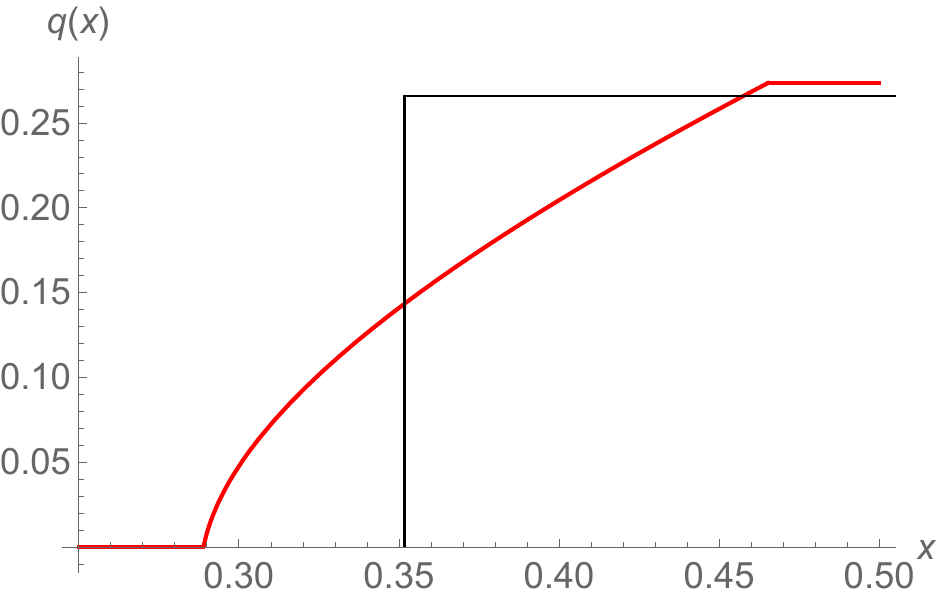}
 \caption{Plots of $q(x)$ for the FRSB solution (red) and the 1RSB solution (black) at $M = 2.50$ for $\tau =-0.01$. This calculation has been done at quintic level, with just one quintic coefficient, with $z_1=z_1^{\rm{eff}}$ and with $y_5=0$, for both the FRSB and 1RSB solutions, in order to simplify the numerical work in the FRSB case. At this value of $M$, the first transition is to the 1RSB state at $\tau=0$, but below the Gardner transition temperature $T_G$, (which corresponds to a value of $\tau_G \approx -0.0078$) there is a transition to a state with FRSB. Below $T_G$, this FRSB state has a higher free energy than the corresponding 1RSB state.}
 \label{Gardnerplot}
 \end{figure}
Physical requirements on the choice of $x_0$ are that for some interval $0<x_1 <  x< x_2< 1$, $q(x)$ is real, an increasing function of $x$, and positive. $x_1$ is for the solutions discussed in this paper at the point where $q(x_1)=0$, and solving this equation gives us $x_1$ as a function of $x_0$. The upper breakpoint, $x_2$, is where  $q(x)$ takes the constant value $q(x_2)$ in the interval $1 > x > x_2$. Its value as a function of $x_0$ is determined by solving Eq.~(\ref{steq:2}) at the value $x = x_2$. This relates the value of $x_2$ to $x_0$. The value of $x_0$ itself can be determined by setting the right-hand side of Eq.~(\ref{eq:D1}) to zero by choosing a value for $x_0$, for any value of $x > x_1$. The FRSB solution for the case $M = 2.25$ at a value of $\tau=-0.001$ is shown in Fig.~\ref{q(x)M2.25}. It is contrasted with the form of $q(x)$ for the 1RSB case at the same values of $M$ and $\tau$.

Note that there is an inverse square root singularity in $q(x)$ when $x=x_s$, where  $y_1-x_s y_3+x_s^2 y_5=0$ but this singularity, the Goldbart-Elderfield singularity, \cite{goldbart:85}, causes no problem so long as it occurs at a value of $x_s$ which is greater than $x_2$ or less than $x_1$. In the limit  $\tau \to 0$, $q(x)$ also goes to zero ($\sim |\tau|$) so Eq.~(\ref{steq:2}) fixes $x_2 \to w_2/w_1 = (M-2)$. Hence a FRSB solution can only exist if $x_2< x_s$, which translates to $M^{**} \leq 2+\sqrt{2}/3 \approx 2.47140$.  The free energy difference between the FRSB and the 1RSB state differs at order $\tau^5$ and we have found numerically that the coefficient of this term goes towards zero as $M \to M^{**}$. One might have thought that one could not ignore the quintic terms when determining $M^{**}$ as they too give a contribution of $O(\tau^5)$. However, in the limit when $\tau \to 0$, both the 1RSB and the FRSB solutions have their upper breakpoints at $w_2/w_1$ and at small $\tau$ the value of $q(x)$ on the plateau is the same for both solutions (see Fig.~\ref{q(x)M2.25}). The form of $q(x)$ for the two solutions only differ in the interval between $x_2$ and $x_1$ and $x_2-x_1 \sim |\tau|$ itself, so in the integrals for the free energy, Eq.~(E1), the plateau regions give the contribution of $O(|\tau|^5)$, which is the same for both solutions, and the region of $x$ where the solutions differ only contributes to the higher order terms in $\tau$.

For $3 > M > M^{**}$ the continuous transition  is to the 1RSB state. For $M >3$, that is  for $w_2/w_1 > 1$, the transition is discontinuous and is to the 1RSB state. We were unable to find a solution with FRSB which had a higher free energy than the 1RSB solution at the discontinuous transition itself.

While the quintic terms are not needed to determine the value of $M^{**}$, it was pointed out years ago that they are needed to obtain the Gardner transition \cite{gross:85}. This is the transition which arises in the 1RSB state and it is to a state with FRSB. Provided we set $y_5$ to zero and just retain one of the quintic terms $z_1$, MATHEMATICA  can analytically solve the first order differential equation, but its explicit form is so long that we have not included its form in this paper. In Fig.~\ref{Gardnerplot} we show the resulting FRSB solution and the 1RSB solution with the same parameters when $M = 2.50$ at a temperature below the Gardner transition temperature, so that the FRSB state has a higher free energy than the 1RSB state. Curiously the form of the FRSB solution is nothing like that given in Ref.~\cite{gross:85}. They claimed that the continuously varying feature of $q(x)$ grew from the upper plateau. However, our solution is very similar to the  FRSB solution for $M < M^{**}$, and it seems natural to us that at low enough temperature that solution should smoothly extend into the region $M > M^{**}$ as $M$ is increased. 

A feature of the Gardner solution is that right at the critical temperature $T_G$ where the Gardner state has a free energy just equal to that of the 1RSB state, its $q(x)$ is such that its derivative  $dq(x)/dx$ is infinite  right at the lower break point $x_1$. This is because at $T_G$ the Goldbart-Elderfield singularity of the quintic order solution is just at $x_1$. As the temperature is reduced below $T_G$, this singularity occurs below $x_1$, and $dq(x)/dx$ is finite at $x_1$ 
(as in Fig.~\ref{Gardnerplot}). For $T > T_G$, the FRSB solution ceases to exist. 

\begin{figure}
 \includegraphics[width=0.9\columnwidth]{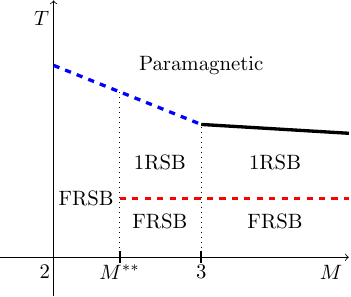}
 \caption{A schematic plot of the phase diagram as a function of $T$ and $M$, within the mean-field approximation. Phase boundaries associated with a continuous transition are drawn with colored dashed lines, while a solid line denotes a discontinuous transition. The FRSB transition for $M > M^{**}$ is the Gardner transition.}
 \label{meanfieldsummaryplot}
 \end{figure}
 
Figure \ref{meanfieldsummaryplot} is a schematic phase diagram showing the phases which we have found in the $M-p$ balanced model as a function of $M$. To find the Gardner phase we had to use the Landau expansion to quintic order. In the next section we shall discuss the effects of the fluctuation corrections to the mean-field theory and argue that in dimensions $d <8$  that the phase diagram becomes radically different to its mean-field form.

 \section{Discussion of Fluctuation Corrections and Behavior in Finite Dimensions}
 \label{sec:discussion}
Most of this paper has been concerned with calculations at mean-field level. Our motivation to study these was because we wished to move towards the inclusion of fluctuations about the mean-field solutions by using RG equations to renormalize the numerous coupling constants, ($\tau$, $w_1$, $w_2$, $y_1$, $\cdots$, $y_5$, $z_1$, $\cdots$, $z_9$) until they lie in the region where fluctuations have become small and mean-field theory becomes accurate. This is the same program as followed by  H\"{o}ller and Read \cite{holler:20}  for the de Almeida-Thouless (AT) transition \cite{almeida:78}. This is the transition of the Ising spin glass in a field $h$, and in the $h-T$ phase diagram there is a line, the de Almeida-Thouless line which separates the high-temperature paramagnetic phase replica symmetric phase from a state with some version of replica symmetry breaking. The field theory of our problem, Eq.~(\ref{quinticLG}) is identical to theirs and the reader should consult their paper for details. However, since their paper was written new simulations have suggested a possible extension of their approach, which we describe.  We begin by briefly summarizing some of their results and procedures.

For the quartic coefficients below $d < 8$ the coefficients $y_1$, $y_2$, $y_3$, $y_4$ and $y_5$ are dominated by the \lq\lq box" diagrams for dimensions $8 > d > 6$ and their bare values become negligible compared to the contribution of the box diagrams, which can be expressed in terms of the values of $w_1$ and $w_2$. For $d > 8$, a good approximation to their values is provided by the bare values of these coefficients.  The important combination of coefficients
\begin{align}
\tilde{y}(x)= Y(x) = y_1 -x y_3 + x^2 y_5,
\label{eq:tildeydef}
\end{align}
 at the value of $x$ corresponding to the upper break point $x_2$ (which in the limit $\tau \to 0$ has the value $w_2/w_1$) plays a key role in determining  the nature of the state below the transition. When $\tilde{y}(\rho)$ is positive,  (where $\rho=w_2/w_1)$), the transition is to a state with FRSB, but  if it is negative the transition is to a state with 1RSB. (This is how the value of $M^{**}$ was determined in the mean-field calculations by setting $x=\rho=M-2$ and solving $Y(x)=0$ for $M$).  H\"{o}ller and Read found from the box diagrams that  
 \begin{align}
 \tilde{y}(\rho)= K_d w_1^4 \rho^2 (22-48 \rho-32 \rho^2 -8 \rho^3 + \rho^4)/(8-d),
 \label{eq:renormy}
 \end{align}
 where $K_d= 2 /(\Gamma(d/2) (4 \pi)^{d/2})$, (provided $\rho <1$).  H\"{o}ller 
 and Read studied in particular the RG flow equations in dimensions $ d = 6 + \epsilon$, where they could employ the Bray and Roberts \cite{bray:80} RG recursion relations. Using these recursion relation, one finds that under the RG transforms $w_1$ and $w_2$ scale down towards zero as $\exp[-\frac{1}{2} \epsilon l]$. As $l\to \infty$ both $w_1$ and $w_2$ approach their fixed point value, (which is $0$) but their ratio $\rho=w_2/w_1$ approaches a constant as the RG scale parameter $l$ goes to infinity. The Bray-Roberts recursion relations are only valid if $w_1$ and $w_2$ are of $O(\sqrt{\epsilon})$ and lie for $d > 6$ within the basin of attraction of the Gaussian fixed point at $w_1=w_2=0$. The bare values of $w_1$ and $w_2$ are of $O(1)$ and so do not lie within the basin of attraction. The fluctuation corrections must somehow first modify the values of $w_1$ and $w_2$ so that the RG calculation can proceed.
 
 It is the numerical value of  $\rho$ in the large $l$ limit which determines whether $\tilde{y}(\rho)$ is positive or negative. The polynomial in Eq.~(\ref{eq:renormy}) is such that $\tilde{y}(\rho)$  is positive provided $\rho < 0.8418$.  H\"{o}ller and Read did not determine the ratio $\rho$.
 We shall argue that its value is universal at least for values of $d < 8$ and that $\rho = 0.5$. Then as $0.5 < 0.8418$, the state formed will have FRSB and so is in the universality class of the Ising spin glass in a field.
 
The key to understanding this is the real space RG calculation of  Angelini and Biroli \cite{angelinibiroli:15}. This suggested that the transition at the AT line in high dimensions might be controlled by a zero-temperature fixed point. They found that in a simple real-space RG approximation that in high enough dimensions, the RG flows of $h$ and $J$, the standard deviation of the bond distribution, which are initially close to their values on the AT line at some non-zero temperature flowed close to their value on the AT line at zero temperature, but then veer away up the $h$-axis at $T=0$.  Then the flow is away from the fixed point at $T= 0$ and $h=h_{AT}$, where $h_{AT}$ is the value of the field $h$ on the AT line at $T=0$. In other words the RG flow is controlled by a zero temperature fixed point. Because their RG procedure  (the Migdal-Kadanoff approximation) works well only in low dimensions it was uncertain whether their zero-temperature fixed point scenario in high dimensions should be trusted. However, we believe that the recent simulation in six dimensions in Ref.~\cite{aguilar:23} strongly suggests that it should be believed. These simulations showed that in six dimensions that the renormalized vertices related to the \lq\lq bare" couplings $w_1$ and $w_2$ were such that their ratio was close to $1/2$. But this is the \textit{same} value (i.e.\ $1/2$) as was found at $T=0$ in the mean-field like Bethe lattice calculation of the same renormalized vertices in Ref.~\cite{rizzo:14}. We therefore shall take it that the renormalized value of $\rho$ which should be inserted into Eq.~(\ref{eq:renormy}) is  $1/2$. As a consequence the continuous transition from the high-temperature phase should be to a state with FRSB, and for $d < 8$ the continuous 1RSB transition should no longer occur.

The same line of argument will also apply to the AT transition of spin glasses in a field. This is a transition from a paramagnetic high-temperature phase to a state with FRSB at lower temperatures. These have been extensively studied by simulations and the most recent of these is that of Bharadwaj et al.\ \cite{bharadwaj:23}. They found numerical evidence  that the AT line might not exist below six dimensions.  The absence of the AT line below six dimensions was argued for in Ref.~\cite{moore:12}, where it was suggested that as $d \to 6$, $h_{AT}^2 \sim (d-6)$, where $h_{AT}$ is the AT field at $T = 0$. If this is correct then in three dimensions there would be no phase transition to a state with replica symmetry breaking, but there could be long length scales according to the droplet picture of spin glasses in a field \cite{mcmillan:84, bray:86, fisher:88} and the Imry-Ma argument \cite{imry:75}, especially if the field is small. That structural glasses might behave as the Ising spin glass in a field was suggested many years ago \cite{moore:06}.

RG calculations are only useful when there exist long correlation length scales. At mean-field level when $\rho=w_2/w_1 > 1$ the transition to the 1RSB state  is via a discontinuous transition at which there are no long correlation length scales. How do the fluctuation corrections affect such a transition? Our belief is that the effect of the fluctuations is to drive the value of the ratio $w_2/w_1$  into the region where the transition is continuous. Certainly there is no sign of a discontinuous transition in the real space RG calculations such as Ref.~\cite{yeo:13}. Nor was there any sign of a discontinuous transition in the AT line simulations in Ref. \cite{bharadwaj:23}. But at present we cannot really exclude the possibility of a discontinuous transition in physical dimensions but  we note once more that the simulations of Ref.~\cite{campellone:98} found no evidence for such a transition at $M =4$ in three dimensions. Fig.~\ref{rgphasessummary} provides a summary of our expected form of the phase diagram first for $6 < d <8$ and secondly for $d <6$.

\begin{figure}
 \includegraphics[width=0.9\columnwidth]{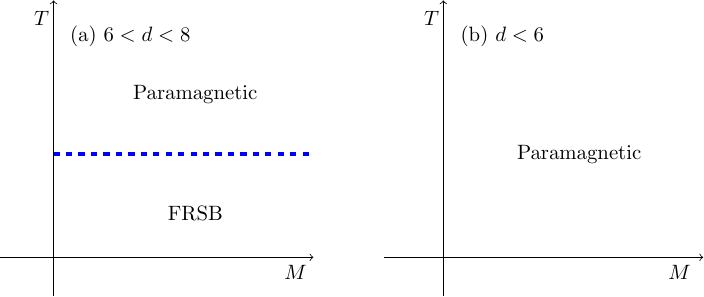}
 \caption{Schematic phase diagrams after allowing for the effect of fluctuation corrections to the mean-field phase diagram of Fig.~\ref{meanfieldsummaryplot} for (a) $6 < d <8$ and (b) $d <6$. For $d <6$ it is hypothesized that there is only one phase present, the high-temperature paramagnetic phase. In the region $6 <d <8$ there is a continuous transition from the paramagnetic phase to a state with FRSB.}
 \label{rgphasessummary}
 \end{figure}

The chief omission of our work is therefore a stronger conclusion on the possible existence of a discontinuous transition and its dependence on the dimensionality $d$ of the system. The only way forward for investigating this question, especially in high dimensions close to or above  $d = 6$ would seem to be simulations on the one-dimensional proxy models. In these proxy models  the form of the long range interactions between the spins can be tuned to mimic behavior in $d$ dimensions. Indeed for the case $p=3$, $M =2$ that has already been done \cite{larson:10}. Alas at mean-field level this model has $w_2/w_1 < 1$ and so it would not be expected to have a discontinuous transition and indeed there was no sign of such in the simulation. The case when $p=3$ and $M =3$ has $w_2/w_1 = 2$ \cite{caltagirone:11} and so might be a good model to simulate as it should have a clear discontinuous transition. The model of the type studied in this paper, $p=4$ but with $M= 4$ could also be a good model to simulate using the one-dimensional proxy model: It has also  $w_2/w_1=2$.
 
\begin{acknowledgments}
We would like to thank Jairo de Almeida for sharing his notes dating from the seventies on the quintic terms in the presence of FRSB for the Ising spin glass. 

\end{acknowledgments}

\appendix

\section{Expansion of the free energy to the quintic order in order parameter}
\label{app:exp}
We expand Eq.~(\ref{L}) to $O(\mu^5)$. We first write $L\equiv 2^{nM}L'$, where
\begin{align}
L'\equiv \mathrm{Tr}'_{\{S^a_i\}} \exp\left[\frac 1{2M} \sum_{(a,b)}\mu_{ab}f_{ab}\right].
\end{align}
Here $\mathrm{Tr}' \equiv 2^{-nM}\mathrm{Tr}$ satisfies $\mathrm{Tr}'_{\{S^a_i\}} 1=1$, and we define
\begin{align}
f_{ab}\equiv \sum_{\alpha=1}^K \Psi^a_\alpha\Psi^b_\alpha ,
\end{align}
where $\bm{\Psi}^a=(S^a_1S^a_2,S^a_1S^a_3,\cdots,S^a_{M-1}S^a_{M})$ is a $K$-dimensional vector
for each replica index $a$ with components $\Psi^a_\alpha$, $\alpha=1,2,\cdots,K\equiv M(M-1)/2$.
The expansion of $L'$ to $O(\mu^5)$ has the following structure:
\begin{align}
&L'= 1+ \tilde{t}_2\sum_{(a,b)}\mu^2_{ab}+
 \tilde{w}_1 \sum_{(a,b,c)}\mu_{ab}\mu_{bc}\mu_{ca}+\tilde{w}_2 \sum_{(a, b)} \mu^3_{ab}  \nonumber \\ 
&+\tilde{y}_1  \sum_{a, b}\mu^4_{ab} 
 +\tilde{y}_2 \sum_{(a,b,c)}\mu^2_{ab}\mu^2_{bc}  
 +\tilde{y}_3 \sum_{(a,b,c)}\mu^2_{ab}\mu_{bc}\mu_{ca}  \nonumber \\ 
 &+\tilde{y}_5  \sum_{(a,b,c,d)}\mu_{ab}\mu_{bc}\mu_{cd}\mu_{da}
 +\tilde{d}_1 \sum_{(a,b,c,d)} \mu^2_{ab} \mu^2_{cd} \nonumber \\
 &+ \tilde{z}_1 \sum_{(a,b)} \mu^5_{ab}  
 +  \tilde{z}_2 \sum_{(a,b,c)} \mu^3_{ab}\mu^2_{bc}
 + \tilde{z}_3 \sum_{(a,b,c)} \mu^3_{ab}\mu_{bc}\mu_{ca} \nonumber \\
 &+ \tilde{z}_4 \sum_{(a,b,c)} \mu^2_{ab}\mu^2_{bc}\mu_{ca}  
 +  \tilde{z}_5 \sum_{(a,b,c,d)} \mu^2_{ab}\mu_{bc}\mu_{cd}\mu_{da} \nonumber\\
 &+  \tilde{z}_6 \sum_{(a,b,c,d)} \mu^2_{ab}\mu_{bc}\mu_{cd}\mu_{db}  
 +\tilde{z}_7 \sum_{(a,b,c,d)} \mu^2_{ab}\mu_{bc}\mu^2_{cd} \nonumber \\
  &+ \tilde{z}_8 \sum_{(a,b,c,d)} \mu_{ab}\mu_{bc}\mu_{cd}\mu_{da}\mu_{ac}  
  +\tilde{z}_9 \sum_{(a,b,c,d,e)} \mu_{ab}\mu_{bc}\mu_{cd}\mu_{de}\mu_{ea}  \nonumber\\
  &+\tilde{d}_2 \sum_{(a,b,c,d)} \mu^3_{ab}\mu^2_{cd}  
  +\tilde{d}_3 \sum_{(a,b,c,d,e)} \mu^2_{ab}\mu_{cd}\mu_{de}\mu_{ec}.
  \label{Lexpansion}
\end{align}
Here $(a,b)$, $(a,b,c)$, $(a,b,c,d)$ etc.\ indicate that the sums  are over all \textit{distinct} replica indices. The coefficients are obtained by taking the trace of the spins
as we explain below.

In order to calculate the free energy, we have to take
the logarithm of $L'$ and expand $\ln(1+x)$ to $O(\mu^5)$. there are three contributions to this order coming from $-(1/2)x^2$ part. They are
\begin{align}
&-\frac 1 2 \tilde{t}^2_2\sum_{(a,b)}\mu^2_{ab}\sum_{(c,d)}\mu^2_{cd}  \label{dis1} \\ 
= &
-\frac 1 2 \tilde{t}^2_2 \left[ 2\sum_{(a,b)}\mu^4_{ab} +4\sum_{(a,b,c)}\mu^2_{ab}\mu^2_{bc}
+\sum_{(a,b,c,d)}\mu^2_{ab}\mu^2_{cd}\right], \nonumber
\end{align}
\begin{align}
&-\frac 1 2 \cdot 2 \tilde{t}_2\tilde{w}_1 \sum_{(a,b)}\mu^2_{ab} \sum_{(c,d,e)}\mu_{cd}\mu_{de}\mu_{ec} \label{dis2}  \\ 
=&- \tilde{t}_2\tilde{w}_1\Big[ 6\sum_{(a,b,c)}\mu^3_{ab}\mu_{bc}\mu_{ca}  
+ 6\sum_{(a,b,c,d)}\mu^2_{ab}\mu_{bc}\mu_{cd}\mu_{db} \nonumber\\
&~~~~~~~~~~ +\sum_{(a,b,c,d,e)} \mu^2_{ab} \mu_{cd}\mu_{de}\mu_{ec} \Big] , \nonumber
\end{align}
and
\begin{align}
&-\frac 1 2 \cdot 2 \tilde{t}_2\tilde{w}_2 \sum_{(a,b)}\mu^2_{ab} \sum_{(c,d)}\mu^3_{cd}  \label{dis3} \\
=&- \tilde{t}_2\tilde{w}_2 \Big[ 2\sum_{(a,b)} \mu^5_{ab} +4\sum_{(a,b,c)}\mu^2_{ab}\mu^3_{bc}  + 
\sum_{(a,b,c,d)}\mu^2_{ab} \mu^3_{cd} \Big] . \nonumber
\end{align}

Note that the last terms in Eqs.~(\ref{dis1}), (\ref{dis2}) and (\ref{dis3}) as well as the terms in Eq.~(\ref{Lexpansion})
with coefficients, $\tilde{d}_i$, $i=1,2,3$ have disconnected parts.
When we take the trace over the spins, we have to keep in mind that the Ising spins must be paired to give nonvanishing contribution. 
For example, we have $\mathrm{Tr}' f_{ab}=0$ for $a\neq b$. 
We evaluate the first few sets of coefficients as follows. 
\begin{align}
&\tilde{t}_2=\frac 1 {2!} \frac{2}{(2M)^2} \; \mathrm{Tr}' f^2_{ab} =\frac 1 {2!} \frac{1}{(2M)^2} 2K =\frac{K}{4M^2},\\
&\tilde{w}_1=\frac 1{3!} \frac{8}{(2M)^3} \; \mathrm{Tr}' f_{ab}f_{bc}f_{ca}=\frac 1{3!} \frac{1}{(2M)^3} 8K 
=\frac{K}{6M^3},\\
&\tilde{w}_2 =\frac 1{3!} \frac{4}{(2M)^3} \; \mathrm{Tr}' f^3_{ab} = \frac 1{3!} \frac{1}{(2M)^3} 4 M(M-1)(M-2) \nonumber \\
&~~~ =\frac{K}{6M^3}(M-2),
\end{align}
and
\begin{align}
\tilde{d}_1&=\frac 1 {4!} \frac{12}{(2M)^4} \; \mathrm{Tr}' f^2_{ab}f^2_{cd} =\frac 1 {4!} \frac{1}{(2M)^4} 12 K^2 ,\\
\tilde{d}_2&=\frac 1{5!} \frac{80}{(2M)^5} \; \mathrm{Tr}' f^3_{ab}f^2_{cd} \nonumber \\
&=\frac 1{5!} \frac{1}{(2M)^5} 80KM(M-1)(M-2),\\
\tilde{d}_3 &=\frac 1{5!} \frac{160}{(2M)^5} \; \mathrm{Tr}' f_{ab}f_{bc}f_{ca} f^2_{de}= \frac 1{5!} \frac{1}{(2M)^5} 160 K^2,
\end{align}
Here
all replica indices are distinct.
One can see that
$\tilde{d}_1= \tilde{t}^2_2/2$, $\tilde{d}_2=\tilde{t}_2\tilde{w}_2$
and
$\tilde{d}_3=\tilde{t}_2\tilde{w}_1$. 
Therefore all the disconnected terms in $\ln L'$ vanish.

\begin{widetext}
We therefore have
\begin{align}
\ln L'=&  ~\tilde{t}_2\sum_{(a,b)}\mu^2_{ab}+
 \tilde{w}_1 \sum_{(a,b,c)}\mu_{ab}\mu_{bc}\mu_{ca}+\tilde{w}_2 \sum_{(a, b)} \mu^3_{ab}  \nonumber \\ 
+&\left( \tilde{y}_1-\tilde{t}^2_2\right)  \sum_{a, b}\mu^4_{ab} 
 +\left( \tilde{y}_2 -2 \tilde{t}^2_2\right) \sum_{(a,b,c)}\mu^2_{ab}\mu^2_{bc}  
 +\tilde{y}_3 \sum_{(a,b,c)}\mu^2_{ab}\mu_{bc}\mu_{ca} 
 +\tilde{y}_5  \sum_{(a,b,c,d)}\mu_{ab}\mu_{bc}\mu_{cd}\mu_{da} \nonumber 
 \\
 +& \left( \tilde{z}_1 - 2 \tilde{t}_2\tilde{w}_2\right)  \sum_{(a,b)} \mu^5_{ab} 
 + \left( \tilde{z}_2 -4 \tilde{t}_2 \tilde{w}_2\right) \sum_{(a,b,c)} \mu^3_{ab}\mu^2_{bc}
 + \left(\tilde{z}_3 -6\tilde{t}_2\tilde{w}_1\right) \sum_{(a,b,c)} \mu^3_{ab}\mu_{bc}\mu_{ca} 
 + \tilde{z}_4 \sum_{(a,b,c)} \mu^2_{ab}\mu^2_{bc}\mu_{ca} \nonumber \\
 +& \tilde{z}_5 \sum_{(a,b,c,d)} \mu^2_{ab}\mu_{bc}\mu_{cd}\mu_{da} 
 +  \left( \tilde{z}_6 -6\tilde{t}_2\tilde{w}_1\right) \sum_{(a,b,c,d)} \mu^2_{ab}\mu_{bc}\mu_{cd}\mu_{db} 
 + \tilde{z}_7 \sum_{(a,b,c,d)} \mu^2_{ab}\mu_{bc}\mu^2_{cd} \nonumber \\
+& \tilde{z}_8 \sum_{(a,b,c,d)} \mu_{ab}\mu_{bc}\mu_{cd}\mu_{da}\mu_{ac} 
  +\tilde{z}_9 \sum_{(a,b,c,d,e)} \mu_{ab}\mu_{bc}\mu_{cd}\mu_{de}\mu_{ea}. \label{lnL_rest}
 \end{align}
 \end{widetext}
 
The first quartic coefficient is given by
\begin{align}
\tilde{y}_1= \frac 1 {4!} \frac{8}{(2M)^4} &  \;\mathrm{Tr}' f^4_{ab} \nonumber \\
= \frac 1 {4!} \frac{8}{(2M)^4} &
 \Big[     K+3K(K-1) \nonumber \\
   &   +3M(M-1)(M-2)(M-3) \Big].
\end{align}
This is valid for $M\ge 3$. For $2\le M\le 3$, there are not enough spins whose combination makes the second term in the square bracket. 
Therefore, the square bracket must be just $K+3K(K-1)$ for $2\le M\le 3$. The rest of them are
\begin{align}
\tilde{y}_2=  \frac 1 {4!} \frac{48}{(2M)^4}   \;\mathrm{Tr}' f^2_{ab}f^2_{bc}= \frac 1 {4!} \frac{48}{(2M)^4} K^2 ,
\end{align}
\begin{align}
\tilde{y}_3=&\frac 1 {4!} \frac{96}{(2M)^4}   \;\mathrm{Tr}' f^2_{ab}f_{bc}f_{ca} \nonumber \\
=&\frac 1 {4!} \frac{96}{(2M)^4} M(M-1)(M-2) ,
\end{align}
and
\begin{align}
\tilde{y}_5=\frac 1 {4!} \frac{48}{(2M)^4} \;\mathrm{Tr}' f_{ab}f_{bc}f_{cd}f_{da}=
\frac 1 {4!} \frac{48}{(2M)^4} K .
\end{align}
These are valid for $M\ge 2$.

We obtain the first quintic coefficient as 
\begin{align}
\tilde{z}_1 =\frac 1 {5!} \frac{16}{(2M)^5} & \; \mathrm{Tr}' f^5_{ab}
\\
=\frac 1 {5!} \frac{16}{(2M)^5} 
    &\Big[  10M(M-1)(M-2)K \nonumber \\
    +&12M(M-1)(M-2)(M-3)(M-4) \Big] .\nonumber 
\end{align}
This is valid for $M\ge 4$. For $2\le M\le 4$, the second term in the square bracket should be dropped for the same reason as
given for $\tilde{y}_1$. The next coefficient is given for $M\ge 2$ as
\begin{align}
\tilde{z}_2 &=\frac 1 {5!} \frac{320}{(2M)^5}  \; \mathrm{Tr}' f^3_{ab}f^2_{bc} \nonumber \\
&=\frac 1 {5!} \frac{320}{(2M)^5} M(M-1)(M-2)K,
\end{align}
The third and fourth quintic coefficients are given by
\begin{align}
\tilde{z}_3=\frac 1 {5!} \frac{320}{(2M)^5} & \; \mathrm{Tr}' f^3_{ab}f_{bc}f_{ca} \nonumber \\
=\frac 1 {5!} \frac{320}{(2M)^5}  &\Big[
      K+3K(K-1) \nonumber \\
      +&3M(M-1)(M-2)(M-3) \Big] , \label{tildez3}
\end{align}
and
\begin{align}
\tilde{z}_4=\frac 1 {5!} \frac{480}{(2M)^5} & \; \mathrm{Tr}' f^2_{ab}f^2_{bc}f_{ca} \nonumber \\
=\frac 1 {5!} \frac{480}{(2M)^5} &\Big[
      2M(M-1)(M-2) \nonumber \\
      +&2M(M-1)(M-2)(M-3) \Big]. \label{tildez4}
\end{align}
Again these expressions are valid only for $M\ge 3$. For $2\le M\le 3$, the second terms in the
square brackets in Eqs.~(\ref{tildez3}) and (\ref{tildez4}) do not appear.
The remaining quintic coefficients are given by 
\begin{align}
\tilde{z}_5 &=\frac 1 {5!} \frac{960}{(2M)^5}  \; \mathrm{Tr}' f^2_{ab}f_{bc}f_{cd}f_{da} \nonumber \\
&=\frac 1 {5!} \frac{960}{(2M)^5} M(M-1)(M-2),
\end{align}
\begin{align}
\tilde{z}_6=\frac 1 {5!} \frac{960}{(2M)^5}  \; \mathrm{Tr}' f^2_{ab}f_{bc}f_{cd}f_{db}
=\frac 1 {5!} \frac{960}{(2M)^5} K^2 ,
\end{align}
\begin{align}
\tilde{z}_7=\frac 1 {5!} \frac{480}{(2M)^5}  \; \mathrm{Tr}' f^2_{ab}f_{bc}f^2_{cd}=0 ,
\end{align}
\begin{align}
\tilde{z}_8 &=\frac 1 {5!} \frac{960}{(2M)^5}  \; \mathrm{Tr}' f_{ab}f_{bc}f_{cd}f_{da}f_{ac} \nonumber \\
&=\frac 1 {5!} \frac{960}{(2M)^5} M(M-1)(M-2),
\end{align}
and
\begin{align}
\tilde{z}_9=\frac 1 {5!} \frac{384}{(2M)^5}  \; \mathrm{Tr}' f_{ab}f_{bc}f_{cd}f_{de}f_{ea}
=\frac 1 {5!} \frac{384}{(2M)^5} K .
\end{align}
These expressions are valid for all $M\ge 2$.

We now convert the summations over replica indices in Eq.~(\ref{lnL_rest}) into those without 
any restriction. We obtain
\begin{align}
&\ln L'=  t'_2\sum_{a,b}\mu^2_{ab}+
 w'_1 \sum_{a,b,c}\mu_{ab}\mu_{bc}\mu_{ca}+w'_2 \sum_{a, b} \mu^3_{ab}  \nonumber \\ 
&+ y'_1 \sum_{a, b}\mu^4_{ab} 
 +y'_2  \sum_{a,b,c}\mu^2_{ab}\mu^2_{bc}  
 +y'_3 \sum_{a,b,c}\mu^2_{ab}\mu_{bc}\mu_{ca} \nonumber \\
 &+y'_5  \sum_{a,b,c,d}\mu_{ab}\mu_{bc}\mu_{cd}\mu_{da} 
  +  z'_1  \sum_{a,b} \mu^5_{ab} 
 +z'_2 \sum_{a,b,c} \mu^3_{ab}\mu^2_{bc} \nonumber \\
 &+ z'_3  \sum_{a,b,c} \mu^3_{ab}\mu_{bc}\mu_{ca} 
 + z'_4 \sum_{a,b,c} \mu^2_{ab}\mu^2_{bc}\mu_{ca} \nonumber \\
& + z'_5 \sum_{a,b,c,d} \mu^2_{ab}\mu_{bc}\mu_{cd}\mu_{da} 
 +  z'_6  \sum_{a,b,c,d} \mu^2_{ab}\mu_{bc}\mu_{cd}\mu_{db} \nonumber \\
& + z'_7 \sum_{a,b,c,d} \mu^2_{ab}\mu_{bc}\mu^2_{cd} 
  + z_8' \sum_{a,b,c,d} \mu_{ab}\mu_{bc}\mu_{cd}\mu_{da}\mu_{ac} \nonumber \\
&  +z'_9 \sum_{a,b,c,d,e} \mu_{ab}\mu_{bc}\mu_{cd}\mu_{de}\mu_{ea},
  \label{lnL}
 \end{align}
where $t'_2=\tilde{t}_2$, $w'_1=\tilde{w}_1$ and $w'_2=\tilde{w}_2$. The first two quartic coefficients are  
\begin{align}
y'_1&=\tilde{y}_1-\tilde{t}^2_2-\left( \tilde{y}_2 -2 \tilde{t}^2_2\right)+\tilde{y}_5  \\
& =\left(\frac{K}{24 M^4} \right)
\begin{cases}
     2 , & \text{if}\ 2\le M \le 3 \\
      (3M^2-15M+20), &\text{if}\ M\ge 3
    \end{cases}
    \nonumber 
\end{align}
and 
\begin{align}
&y'_2= \tilde{y}_2 -2 \tilde{t}^2_2 -2 \tilde{y}_5 =-\left(\frac{K}{4M^4}\right) .
\end{align}
The rest of them are the same as when the summations are restricted.
\begin{align}
y'_3=\tilde{y}_3,~~~ 
y'_5=\tilde{y}_5.
\end{align}

The quintic coefficients are given by
\begin{align}
z'_1&=\tilde{z}_1 - 2 \tilde{t}_2\tilde{w}_2 
-\left( \tilde{z}_2 -4 \tilde{t}_2 \tilde{w}_2\right)+\tilde{z}_5+\tilde{z}_7 \\
&=\left(\frac{K}{10M^5}\right) \begin{cases}
    5(M-2) , & \text{if}\ 2\le M \le 4 \\
      (M-2)(M^2-7M+17), &\text{if}\ M\ge 4
    \end{cases}
    \nonumber 
    \end{align}
 \begin{align}
z'_2 &=\tilde{z}_2 -4 \tilde{t}_2 \tilde{w}_2-2\tilde{z}_5-2\tilde{z}_7 \nonumber \\
&=-\left(\frac{K}{M^5}\right)(M-2),
\end{align}
\begin{align}
z'_3 &=\tilde{z}_3 -6\tilde{t}_2\tilde{w}_1-2\left( \tilde{z}_6 -6\tilde{t}_2\tilde{w}_1\right)+5\tilde{z}_9 \\
& =\left(\frac{K}{6M^5}\right) 
\begin{cases}
     2 , & \text{if}\ 2\le M \le 3 \\
     (3M^2-15M+20), &\text{if}\ M\ge 3
\end{cases}
\nonumber 
\end{align}
\begin{align}
z'_4 &=\tilde{z}_4-\tilde{z}_7-\tilde{z}_8 \\
&=\left(\frac{K}{2M^5}\right)
\begin{cases}
    0 , & \text{if}\ 2\le M \le 3 \\
      (M-2)(M-3), &\text{if}\ M\ge 3
\end{cases}
\nonumber 
\end{align}
and 
\begin{align}
z'_6=\tilde{z}_6 -6\tilde{t}_2\tilde{w}_1-5\tilde{z}_9=
-\left(\frac{K}{2M^5}\right).
\end{align}
The other coefficients are unchanged, namely,
\begin{align}
z'_i = \tilde{z}_i
\end{align}
for $i=5,7,8$ and $9$.

Finally, the free energy is now given by Eq.~(\ref{free1}) with Eq.~(\ref{G}). One of the saddle point equations gives
$\mu_{ab}=\beta^2 q_{ab}$. Inserting this relation into Eq.~(\ref{free1}), we obtain the free energy in the form given in Eq.~(\ref{quinticLG})
with 
\begin{align}
w_i\equiv\beta^6 w'_i, ~~~~~ y_j\equiv \beta^8 y'_j,~~~~~ z_k\equiv \beta^{10} z'_k ,
\end{align}
for $i=1,2$, $j=1,2,3,5$ and $k=1,2,\cdots,9$.

\section{Small-$\sigma$ behavior of $f_M(\sigma)$}
\label{app:smalls}

Here we present some steps leading to the small-$\sigma$ expansion of $f_M(\sigma)$ defined in Eq.~(\ref{eq_m}).
As mentioned in the main text, we expand $f_M(\sigma)$ up to $O(\sigma^{8})$. There are numerous terms
to be evaluated. In the following, for brevity, we only list the quantities needed for the calculation of the $O(\sigma^6)$-coefficient.
We first write
\begin{align}
\zeta(\bm{y},\mu_1)\equiv\frac 1 {2^M}\underset{\{S_i\}}{\mathrm{Tr}}\;  \exp\left[ \sigma\bm{y}\cdot\bm{\Psi}
\right]=\sum_{j=0}^\infty \frac{\sigma^j}{j!} \zeta_j(\bm{y}),
\end{align}
where $\sigma\equiv\sqrt{\mu_1/M}$.
We immediately see that $\zeta_1(\bm{y})=0$ since $\mathrm{Tr}\;\Psi_\alpha=0$. Using the fact that $\mathrm{Tr}\Psi_\alpha\Psi_\beta=0$ for $\alpha\neq \beta$, we find that $\zeta_2(\bm{y})=\sum_{\alpha}^K y^2_\alpha$ and
$\zeta_3(\bm{y})=\sum^K_{ (\alpha,\beta,\gamma)}y_\alpha y_\beta y_\gamma\; \frac 1 {2^M}\mathrm{Tr}\Psi_\alpha\Psi_\beta\Psi_\gamma $. Higher order contributions are 
\begin{align}
\zeta_4(\bm{y})&=\sum_\alpha^K y^4_\alpha 
+3\sum^K_{\alpha\neq \beta} y^2_\alpha y^2_\beta \nonumber \\
&+\sum^K_{ (\alpha,\beta,\gamma,\delta)}y_\alpha y_\beta y_\gamma y_\delta \;\frac 1 {2^M}\mathrm{Tr}\Psi_\alpha\Psi_\beta\Psi_\gamma
\Psi_\delta,
\end{align}
\begin{align}
\zeta_5(\bm{y})&=10\sum^K_{ (\alpha,\beta,\gamma)}y^3_\alpha y_\beta y_\gamma \;  \frac 1 {2^M}\mathrm{Tr}\Psi_\alpha\Psi_\beta\Psi_\gamma  
\nonumber \\
&+10\sum^K_{ (\alpha,\beta,\gamma,\delta)}y^2_\alpha y_\beta y_\gamma y_\delta\; 
\frac 1 {2^M}\mathrm{Tr}\Psi_\beta\Psi_\gamma\Psi_\delta  \\
&+\sum^K_{ (\alpha,\beta,\gamma,\delta,\sigma)}y_\alpha y_\beta y_\gamma y_\delta y_\sigma \; \frac 1 {2^M}\mathrm{Tr}\Psi_\alpha\Psi_\beta\Psi_\gamma \Psi_\delta\Psi_\sigma, \nonumber 
\end{align}
and
\begin{align}
\zeta_6(\bm{y})&=\sum_\alpha^K y^6_\alpha +15\sum^K_{\alpha\neq \beta} y^4_\alpha y^2_\beta 
+15 \sum^K_{ (\alpha,\beta,\gamma)}y^2_\alpha y^2_\beta y^2_\gamma  \\
&+20\sum^K_{ (\alpha,\beta,\gamma,\delta)}y^3_\alpha y_\beta y_\gamma y_\delta \;\frac 1 {2^M}\mathrm{Tr}\Psi_\alpha\Psi_\beta\Psi_\gamma
\Psi_\delta\nonumber \\
&+15 \sum^K_{ (\alpha,\beta,\gamma,\delta,\sigma)}y^2_\alpha y_\beta y_\gamma y_\delta y_\sigma \; \frac 1 {2^M}\mathrm{Tr}\Psi_\beta\Psi_\gamma \Psi_\delta\Psi_\sigma \nonumber \\
&+\sum^K_{ (\alpha,\beta,\gamma,\delta,\sigma,\mu)}y_\alpha y_\beta y_\gamma y_\delta y_\sigma y_\mu\; \frac 1 {2^M}\mathrm{Tr}\Psi_\alpha\Psi_\beta\Psi_\gamma \Psi_\delta\Psi_\sigma\Psi_\mu . \nonumber 
\end{align}
Here $(\alpha,\beta,\gamma)$, etc indicate the summation is over all distinct indices
and $K\equiv {M \choose 2}$. 
Performing the Gaussian integrals, we have $\int D^K \bm{y}\; \zeta_j(\bm{y})=0$ for $j$ odd, 
$\int D^K \bm{y}\; \zeta_2(\bm{y}) = K $, $\int D^K \bm{y}\; \zeta_4(\bm{y}) =  3K +3K(K-1) $, and
\begin{align}
\int D^K \bm{y}\; \zeta_6(\bm{y}) &= 15K+45K(K-1) \nonumber \\
& +15K(K-1)(K-2) . \label{int1}
\end{align}
For the calculation up to $O(\sigma^6)$, we also need the following quantities:
 \begin{align}
 \int D^K \bm{y}\; \zeta^2_2(\bm{y}) =& 3K+ K(K-1),\\
 \int D^K \bm{y}\; \zeta^3_2(\bm{y}) =& 15K+9K(K-1) \nonumber \\
 &+K(K-1)(K-2) ,\\
 \int D^K \bm{y}\; \zeta_2(\bm{y})\zeta_4(\bm{y}) =& 15K+21K(K-1) \nonumber \\
& +3K(K-1)(K-2) ,\\
  \int D^K \bm{y}\; \zeta^2_3(\bm{y}) = & 6M(M-1)(M-2) . \label{int2}
 \end{align}
 These expressions are valid when $K\ge 2$ or $M=3,4,5,\cdots$.

Now in the second term inside the integral in Eq.~(\ref{eq_m}), we can write by symmetry
\begin{align}
&\sum_{\alpha=1}^K\left[ \frac 1{2^{M}}\mathrm{Tr}\; \Psi_ \alpha \exp\left[ \sigma\bm{y}\cdot\bm{\Psi}\right]\right]^2
\nonumber \\
=&
K\left[ \frac 1{2^{M}}\mathrm{Tr}\; \Psi_1 \exp\left[ \sigma\bm{y}\cdot\bm{\Psi}\right]\right]^2.
\end{align}
We then define
\begin{align}
\frac 1{2^{M}}\mathrm{Tr}\; \Psi_ 1 \exp\left[ \sigma\bm{y}\cdot\bm{\Psi}\right]\equiv \sum_{j=1}^{\infty} \frac{\sigma^j}{j!}\eta_j(\bm{y}).
\end{align}
We find that $\eta_1(\bm{y})=y_1$, 
\begin{align}
\eta_2(\bm{y})=\sum_{(\alpha,\beta)} y_\alpha y_\beta 2^{-M}\mathrm{Tr}\Psi_1\Psi_\alpha\Psi_\beta,
\end{align}
and
\begin{align}
\eta_3(\bm{y})&= y_1+3y_1\sum_{\alpha\neq 1} y^2_\alpha \nonumber \\
&+\sum^K_{ (\alpha,\beta,\gamma)}y_\alpha y_\beta y_\gamma\; \frac 1 {2^M}\mathrm{Tr}\Psi_1\Psi_\alpha\Psi_\beta\Psi_\gamma .
\end{align}
For the calculation up to $O(\sigma^6)$, we need
\begin{align}
&\int D^K \bm{y}\; \eta^2_2(\bm{y}) = 4(M-2) ,\label{int3}\\
&\int D^K \bm{y}\; \eta_1(\bm{y}) \eta_3(\bm{y}) = 3K ,\\
&\int D^K \bm{y}\; \eta^2_1(\bm{y}) \zeta_2(\bm{y})=K+2 . \label{int4}
\end{align}

It is now a matter of Taylor expanding the functions inside the integral in Eq.~(\ref{eq_m}) and 
using the above results to get the expansion coefficients in 
$f_M(\sigma)=\sum_{j=0}^\infty c_{2j}(M) \sigma^{2j}$.
We find that $c_0=c_2=c_4=0$ and the leading order term is $O(\sigma^6)$. 
We obtain
\begin{align}
c_6(M)=-\frac M{24} (M-1)(M-3).
\label{result}
\end{align}
As mentioned in the main text, it becomes negative for $M>3$.
To go up to $O(\sigma^{8})$, we need results of more Gaussian integrals similar to Eqs.~(\ref{int1})-(\ref{int2}) and
to Eqs.~(\ref{int3})-(\ref{int4}).
After a rather long calculation with the help of symbolic algebra packages in MATHEMATICA, we obtain
\begin{align}
c_8(M)=-\frac{M}{48} (M-1)(3M^2-27M+47),
\end{align}
which is valid for $K\ge 3$ or $M=3,4,5,\cdots$.
We note that $c_8(M=3)=7/8>0$.

\section{The 1RSB equations for the quintic Landau free energy}
\label{app:1rsb}
Here we consider the 1RSB saddle point equations corresponding to the free energy expanded up to quintic order as given in Eq.~(\ref{quinticLG}).
Let us assume that $q_{ab}$ takes the 1RSB form 
having values $q_1$ on $n/m_1$ diagonal blocks 
of size $m_1$ and 
$q_0=0$ outside the blocks. We can then express the cubic, quartic, and quintic terms in $q_{ab}$ in terms of $q_1$ and $m_1$
as we have done in Eqs.~(\ref{qp}) and (\ref{lamp}) for the quadratic terms. We obtain
\begin{widetext}
\begin{align}
\frac{\beta F_{\rm 1RSB}}{N}=&-C\beta^2-M\ln 2+\tau (m_1-1)q_1^2 
-w_1(m_1-1)(m_1-2)q^3_1 - w_2 (m_1-1)q^3_1   \nonumber \\
&- y_1 (m_1-1) q^4_1 
-y_2 (m_1-1)^2 q^4_1
- y_3(m_1-1)(m_1-2)q^4_1  -y_5 (m_1-1)
(m_1^2-3m_1+3)q^4_1  \nonumber \\
&-z_1 (m_1-1) q^5_1 -z_2(m_1-1)^2 q^5_1 -z_3(m_1-1)(m_1-2) q^5_1 -z_4(m_1-1)(m_1-2) q^5_1 \nonumber \\
&-z_5 (m_1-1)(m_1^2-3m_1+3) q^5_1-z_6 (m_1-1)^2(m_1-2) q^5_1 -z_7(m_1-1)^3 q_1^5 \nonumber \\
&-z_8 (m_1-1)(m_1-2)^2 q^5_1 -z_9 (m_1-1)(m_1-2)(m_1^2-2m_1+2)q^5_1 . \label{LGfree1rsb}
\end{align}
\end{widetext}
The saddle point equations are obtained by varying the free energy with
respect to $q_1$ and $m_1$. They are given by 
\begin{align}
2 \tau q_1 = & 3 \Big[ w_1 (m_1-2) +w_2 \Big] q^2_1 
+ 4 \Big[ y_1+y_2(m_1-1) \nonumber \\
&+y_3(m_1-2)+y_5 (m_1^2-3m_1+3) \Big] q^3_1 \nonumber \\
+ &5\Big[  z_1   + z_2(m_1-1)  + z_3 (m_1-2)  + z_4 (m_1-2)  \nonumber \\
&+z_5 (m_1^2-3m_1+3) 
 + z_6 (m_1-1)(m_1-2)  \nonumber \\
& + z_7(m_1-1)^2  
+z_8 (m_1-2)^2  \nonumber \\
&+z_9 (m_1-2)(m_1^2-2m_1+2) \Big] q^4_1
\label{q1_gen}
\end{align}
and
\begin{align}
\tau q^2_1 =&  \Big[ w_1 (2 m_1-3) +w_2 \Big] q^3_1 
+  \Big[ y_1+ 2y_2(m_1-1) \nonumber \\
& +y_3( 2m_1-3)+y_5 (3m^2_1-8m_1+6) \Big] q^4_1 \nonumber \\
+& \Big[  z_1   + 2 z_2(m_1-1)  + z_3 (2m_1-3)  + z_4 (2m_1-3)  \nonumber \\
&+z_5 (3m^2_1-8 m_1 +6) + z_6 (3m^2_1-8m_1+5) \nonumber \\
& + 3 z_7(m_1-1)^2  
+z_8 (3m^2_1-10m_1+8)  \nonumber \\
&+z_9 (4m^3_1-15m^2_1+20 m_1-10) \Big] q^5_1
\label{m1_gen}
\end{align}
Combining the above equations with the condition $q_1\neq 0$, we have
\begin{align}
0 =&  \Big[ -m_1 w_1  +w_2 \Big] 
+  2 \Big[ y_1 - y_3 +y_5 m_1(2-m_1) \Big] q_1 \nonumber \\
 + &\Big[  3 z_1   +  z_2(m_1-1)  + z_3 (m_1-4)  + z_4 (m_1-4)  \nonumber \\
+&z_5 (-m^2_1+ m_1 +3)  + z_6 m_1(1-m_1) - z_7(m_1-1)^2  \nonumber \\
+ &z_8 (4-m^2_1)  +z_9 m_1(-3 m^2_1+10m_1-10) \Big] q^2_1
\label{1rsb_gen}
\end{align}

The 1RSB transition temperature is determined by setting $m_1=1$ in the above equation. 
We obtain
\begin{align}
&(w_2-w_1) +2 (y_1-y_3+y_5)q_1 \nonumber \\
&+3(z_1-z_3-z_4+z_5+z_8-z_9)q^2_1=0.
\label{1rsb_exp}
\end{align}
Equivalently, we have an equation without factors of $\beta$ as
\begin{align}
&(w'_2-w'_1) +2 (y'_1-y'_3+y'_5)\mu_1 \nonumber \\
&+3(z'_1-z'_3-z'_4+z'_5+z'_8-z'_9)\mu^2_1=0.
\label{1rsb_exp1}
\end{align}
From Appendix \ref{app:exp}, the coefficients are given by
\begin{align}
w_2-w_1= \frac{\beta^6}{12 M^2}(M-1)(M-3),
\end{align}
and
\begin{align}
&y_1-y_3+y_5 \\
=&
\begin{cases}
      -\frac{\beta^{8}}{48 M^3}(M-1) (12M-29), & \text{if}\ 2\le M \le 3 \\
     \frac{\beta^{8}}{48 M^3}(M-1)(3M^2-27M+47), &\text{if}\ M\ge 3 .
\end{cases}
\nonumber 
\end{align}
In Sec.~\ref{sec:frsb}, we have defined the effective quintic coefficient $z_1^{\rm eff}$ as the one that appears 
in the above equation, which can be calculated from the results in Appendix \ref{app:exp} as
\begin{align}
z_1^{\rm eff}&\equiv z_1-z_3-z_4+z_5+z_8-z_9  \\
&=
\begin{cases}
      \frac{\beta^{10}}{60 M^4}(M-1) (45M-103), 
       \\
     -\frac{\beta^{10}}{60 M^4}(M-1)(30M^2-195M+283), 
     \\
     \frac{\beta^{10}}{60 M^4}(M-1)(3 M^3-57 M^2+273 M -355), 
\end{cases}
\nonumber
\label{app:z1eff}
\end{align}
In the above equation, the three cases from top to bottom correspond to the regions, $2\le M\le 3$, $3\le M\le 4$ and
$M\ge 4$, respectively.
This is related to the small-$\sigma$ expansion of $f_M(\sigma)$ discussed in Sec.~\ref{sec:1rsb} as follows.
If we multiply Eq.~(\ref{1rsb_exp}) by $-q^3_1/2$ and use $q_1=\mu_1/\beta^2 =M\sigma^2/\beta^2$, Eq.~(\ref{1rsb_exp})
becomes
\begin{align}
c_6(M) \sigma^6 + c_8(M)\sigma^8+c_{10}(M)\sigma^{10}=0,
\end{align}
where
\begin{align}
&c_6(M)=-\frac{M^3}{2\beta^6}(w_2-w_1), \\
&c_8(M)= -\frac{M^4}{\beta^8} (y_1-y_3+y_5), 
\end{align}
and
\begin{align}
&c_{10}(M)=-\frac{3M^5}{2\beta^{10}} (z_1-z_3-z_4+z_5+z_8-z_9). \label{app:c10}
\end{align}

\section{FRSB equations for the free energy with one quintic term}
\label{app:frsbeq}

Taking a functional derivative of the free energy in Eq.~(\ref{FRSB}) with respect to $q(x)$, we have
\begin{widetext}
\begin{align}
0=&\frac\delta{\delta q(x)} \left( \frac{\beta F_{\rm FRSB}}{N} \right) = -2\tau q(x) -w_1\left\{ 3x q^2(x) +3\int_0^x dy\; q^2(y) +6q(x)\int_x^1 dy\; q(y)\right\}
+3 w_2 q^2(x) \nonumber \\
&+4y_1 q^3(x) -4 y_2   \langle q^2 \rangle q(x) 
-y_3\left\{ 2 \langle q^3\rangle +6 \langle q\rangle q^2(x) 
+2 \langle q^2 \rangle  q(x) +4 x q^3(x) -6q^2(x) \int_0^x dy\; q(y) -2 \int_x^1 dy q^3(y) \right\}
\nonumber \\
&-y_5\Bigg\{ 4\langle q^2 \rangle q(x) -8 \langle q \rangle^2 q(x) -8\langle q\rangle \langle q^2 \rangle -4\int_0^1 dx'\; q(x')\int_0^{x'} dy\; (q(x')-q(y))^2 
\nonumber \\
&~~~ -4\langle q\rangle \Big[ 3x q^2(x)-4q(x)\int_0^x dy\; q(y)-2\int_x^1 dy\; q^2(y)+\int_0^x dy\; q^2(y) +2q(x)\int_x^1 dy\; q(y)\Big] \nonumber \\
&~~~ -\Bigg[ 4x^2 q^3(x) -12 x q^2(x) \int_0^x dy\; q(y) -4\int_x^1 dy\; y q^3(y) +4x q(x)\int_0^x dy\; q^2(y) + 4q(x) \int_x^1 dy\; y q^2(y) \nonumber \\
&~~~~~~  -4\int_0^x dy q(y) \int_0^x dz\;q^2(z) -4\int_x^1 dy\; q(y) \int_0^y dz\; q^2(z) 
 - 8q(x) \int_x^1 dy\; q(y)\int_0^y dz\; q(z) +8 q(x)\left[ \int_0^x dy\; q(y) \right]^2   \nonumber \\
&~~~~~~  +8 \int_x^1 dy\; q^2(y) \int_0^y dz\; q(z) +4 q(x)\int_x^1 dy\; \int_0^y dz\; q^2(z) \Bigg] \Bigg\} +5 z_1 q^4(x).
\label{eq:D1}
\end{align}
\end{widetext}

For $0\le x \le 1$ where $q'(x)\neq 0$, we can take a derivative of the above equation and have
\begin{align}
 \left(\frac{1}{q'(x)}\frac{d}{dx} \right) \left[ \frac\delta{\delta q(x)} \left( \frac{\beta F_{\rm FRSB}}{N} \right) \right] =0.
 \label{eq:0}
\end{align}
This gives us
\begin{align}
0=& -2\tau -w_1\left\{ 6x q(x) +6 \int_x^1 dy\; q(y)\right\} +6w_2 q(x) \nonumber \\
+&12 y_1 q^2(x)-4 y_2 \langle q^2 \rangle   
-y_3 \Bigg\{ 12 \langle q \rangle q(x)+12 x q^2(x) \nonumber \\
& -12 q(x)\int_0^x dy\; q(y) +2\langle q^2 \rangle\Bigg\} 
-y_5 \Bigg\{ 4 \langle q^2 \rangle -8 \langle q\rangle^2 \nonumber \\
& -4\langle q\rangle 
\Big[6x q(x)-4\int_0^x dy\; q(y) +2\int_x^1 dy\; q(y)\Big] \nonumber \\
&-\Bigg[ 12x^2 q^2(x)-24xq(x)\int_0^x dy\; q(y)+4x \int_0^x dy\; q^2(y) \nonumber \\
&~~ +4\int_x^1 dy\; y q^2(y) -8\int_x^1 dy\; q(y)\int_0^y dz\; q(z)\nonumber \\
&~~  +8 \left[ \int_0^x dy q(y) \right]^2 +4\int_x^1 dy\int_0^y dz\; q^2(z)\Bigg]\Bigg\} \nonumber \\
+&20 z_1 q^3(x)
\end{align}
Taking one more derivative with respect to $x$ and divide by $q'(x)$, we have for $x$ with $q'(x)\neq 0$
\begin{align}
 \left( \frac{1}{q'(x)}\frac{d}{dx} \right) \left(\frac{1}{q'(x)}\frac{d}{dx} \right) \left[ \frac\delta{\delta q(x)} \left( \frac{\beta F_{\rm FRSB}}{N} \right) \right] =0.
 \label{eq:1}
\end{align}
This is given by
\begin{align}
0
= &-6(w_1 x- w_2) +24 Y(x) q(x) \nonumber \\
&+12 Y'(x)  \int_x^1 dy\; q(y)  +60 z_1 q^2(x),
\label{steq:1}
\end{align}
where
\begin{align} 
Y(x)\equiv y_1 -x y_3 +x^2 y_5.
\end{align}
Taking a derivative of the above equation with respect to $x$ once again, we have 
\begin{align}
\frac{d}{dx} \left( \frac{1}{q'(x)}\frac{d}{dx} \right) \left(\frac{1}{q'(x)}\frac{d}{dx} \right) \left[ \frac\delta{\delta q(x)} \left( \frac{\beta F_{\rm FRSB}}{N} \right) \right] =0,
\label{eq:2}
\end{align}
This can be written as
\begin{align}
0
=& -6 w_1 +24 Y(x) q'(x) +12 Y'(x) q(x) \nonumber \\
&+24 y_5 \int_x^1 dy\; q(y)  +120 z_1 q'(x) q(x).
\label{steq:2}
\end{align}
Eliminating $\int_{x_0}^1 dy\; q(y)$ from Eqs.~(\ref{steq:1}) and (\ref{steq:2}), we have
\begin{align}
&q'(x) \nonumber \\
=& \frac{-y_3 w_1 +2 y_5 w_2+2 (-y^2_3+4 y_1 y_5)q(x)+20  z_1 y_5 q^2(x)}
{4Y'(x)(Y(x)+5 z_1 q(x) )}.  
\label{qprime0}
\end{align}


\section{FRSB expressions for all quintic terms}
\label{app:frsbquintic}

Here we present the expressions in terms of the Parisi function $q(x)$ for the quintic contributions to the free energy,
which is denoted by $F^{(5)}_{\rm FRSB}$.
We have
\begin{widetext}
\begin{align}
\frac{\beta F_{\rm FRSB}^{(5)}}{N}&= z_1 \langle q^5 \rangle
-z_2 \Big[ -\langle q^5 \rangle +2\langle q^3\rangle \langle q^2\rangle +\int_0^1 dx\; \int_0^x dy\; ( q^3(y)-q^3(x))(q^2(y)-q^2(x))\Big] \nonumber \\
&-z_3 \left[ 2\langle q\rangle \langle q^4\rangle +\int_0^1 dx\; q^3(x) \int_0^x dy\; (q(y)-q(x))^2 \right] 
-z_4 \left[ 2\langle q^2\rangle \langle q^3\rangle +\int_0^1 dx\; q(x) \int_0^x dy\; \left( q^2(y)-q^2(x)\right)^2\right] \nonumber \\
&-z_5 \Big [ -4\langle q\rangle^2 \langle q^3\rangle +\langle q^2 \rangle \langle q^3 \rangle -3\langle q\rangle  \langle q^2  h \rangle 
-\langle q^3 \rangle \langle h\rangle  -\int_0^1 dx\; q^2(x)  \int_0^x dy\; (q(y)-q(x))(h(y)-h(x)) \Big], \nonumber \\
&-z_6 \left[ -2\langle q \rangle \langle q^2\rangle ^2 -\langle q^2\rangle  \langle q h\rangle \right]
-z_7 \Big[  2\langle q^2\rangle \langle q^3\rangle +\langle q \rangle \langle q^4\rangle -4\langle q\rangle \langle q^2\rangle ^2
-3\langle q^2 \rangle \langle g\rangle 
+\langle q^2 g\rangle \nonumber \\
&~~~~~~~~ -\langle q\rangle \int_0^1 dx \int_0^x dy\; \left( q^2(y)-q^2(x)\right)^2 
-\int_0^1 dx\int_0^x dy\; (g(y)-g(x))(q^2(y)-q^2(x))\Big], \nonumber \\
&-z_8 \left[ -4\langle q^2\rangle \langle q^3\rangle -4\langle q\rangle \langle q^2 h\rangle -\langle q h^2\rangle \right] 
-z_9\Big[ 8\langle q \rangle^3\langle q^2 \rangle-4\langle q^2 \rangle^2 \langle q \rangle+10 \langle q \rangle^2 \langle q h \rangle
-2 \langle q^2 \rangle \langle q h \rangle \nonumber \\
&~~~~~~~ +2\langle q \rangle\langle q^2 \rangle\langle h \rangle +3\langle q \rangle\langle h^2 \rangle
+\langle h \rangle\langle q h\rangle 
+2\langle q \rangle\int_0^1 dx\; q(x)\int_0^x dy\; (q(y)-q(x))(h(y)-h(x)) \nonumber \\
&~~~~~~~ +
\int_0^1 dx\; h(x)\int_0^x dy\; dy (q(y)-q(x))(h(y)-h(x)) \Big],
\end{align}
\end{widetext}
where
\begin{align}
&h(x)=\int_0^x dy\; (q(y)-q(x))^2  \\
&g(x)=\int_0^x dy\; \left( q^2(y)-q^2(x)\right)(q(y)-q(x))
\end{align}

Stationary conditions for the free energy obtained from the quintic contributions are quite complicated. In this Appendix, we only present 
\begin{align}
0=\left( \frac{1}{q'(x)}\frac{d}{dx} \right) \left(\frac{1}{q'(x)}\frac{d}{dx} \right) \left[ \frac\delta{\delta q(x)} \left( \beta F_{\rm FRSB}^{(5)}/N\right) \right].
\end{align}
This is given by
\begin{widetext}
\begin{align}
0&= 60 z_1 q^2(x)- 6 z_2 \langle q^2 \rangle -z_3 \left[ 6\int_0^x dy\; q^2(y)+ 48 q(x) \int_x^1 dy\; q(y) + 60 x q^2(x) \right] \nonumber \\
&-z_4\left[ 12\int_x^1 dy\; q^2(y) +24q(x) \int_x^1 dy\; q(y) +60x q^2(x) \right] 
-z_5\Big[  -24\langle q\rangle^2+6\langle q^2\rangle-6\langle h\rangle -72 \langle q\rangle  xq(x) \nonumber \\
&~~~~~~~ + 36 \langle q\rangle \int_0^x dy\; q(y) 
-6 x \int_x^1 dy\; q^2(y) +60 x q(x) \int_0^x dy\; q(y) -12 \left(\int_0^x dy\; q(y)\right)^2 \nonumber \\
&~~~~~~~ -54 x^2 q^2(x) +6\int_0^x dy\; h(y) -6x h(x)\Big] \nonumber \\
&-z_6 \left[-6\langle q^2 \rangle x \right] 
-z_8 \Big[ -24\langle q^2\rangle -96 \langle q \rangle x q(x)+48 \langle q\rangle \int_0^x dy\; q(y)
+96 x q(x)\int_0^x dy\; q(y) \nonumber \\
&~~~~~~~-24 \left( \int_0^x dy\; q(y)\right)^2  -60 x^2 q^2(x)-12 x \int_0^x dy\; q^2(y) \Big] \nonumber \\
&-z_9\Big[ 12 x \langle q \rangle^2 +48 x \langle q \rangle \{ x q(x)-\int_0^x dy\; q(y)\} 
+12 x^2 h(x) + 6x \left(\langle h \rangle -2 \int_0^x dy\; h(y)\right) \nonumber \\
&~~~~~~~ +48 x \left(x q(x)-\int_0^x dy\; q(y)\right)^2  +6 x \langle h \rangle
+72 x \langle q \rangle \int_0^x dy\; (q(x)-q(y))-12 x \langle q^2 \rangle +60 x \langle q \rangle^2 \Big] .
\end{align}
\end{widetext}





\bibliography{refs}

\begin{thebibliography}{34}%
\makeatletter
\providecommand \@ifxundefined [1]{%
 \@ifx{#1\undefined}
}%
\providecommand \@ifnum [1]{%
 \ifnum #1\expandafter \@firstoftwo
 \else \expandafter \@secondoftwo
 \fi
}%
\providecommand \@ifx [1]{%
 \ifx #1\expandafter \@firstoftwo
 \else \expandafter \@secondoftwo
 \fi
}%
\providecommand \natexlab [1]{#1}%
\providecommand \enquote  [1]{``#1''}%
\providecommand \bibnamefont  [1]{#1}%
\providecommand \bibfnamefont [1]{#1}%
\providecommand \citenamefont [1]{#1}%
\providecommand \href@noop [0]{\@secondoftwo}%
\providecommand \href [0]{\begingroup \@sanitize@url \@href}%
\providecommand \@href[1]{\@@startlink{#1}\@@href}%
\providecommand \@@href[1]{\endgroup#1\@@endlink}%
\providecommand \@sanitize@url [0]{\catcode `\\12\catcode `\$12\catcode
  `\&12\catcode `\#12\catcode `\^12\catcode `\_12\catcode `\%12\relax}%
\providecommand \@@startlink[1]{}%
\providecommand \@@endlink[0]{}%
\providecommand \url  [0]{\begingroup\@sanitize@url \@url }%
\providecommand \@url [1]{\endgroup\@href {#1}{\urlprefix }}%
\providecommand \urlprefix  [0]{URL }%
\providecommand \Eprint [0]{\href }%
\providecommand \doibase [0]{http://dx.doi.org/}%
\providecommand \selectlanguage [0]{\@gobble}%
\providecommand \bibinfo  [0]{\@secondoftwo}%
\providecommand \bibfield  [0]{\@secondoftwo}%
\providecommand \translation [1]{[#1]}%
\providecommand \BibitemOpen [0]{}%
\providecommand \bibitemStop [0]{}%
\providecommand \bibitemNoStop [0]{.\EOS\space}%
\providecommand \EOS [0]{\spacefactor3000\relax}%
\providecommand \BibitemShut  [1]{\csname bibitem#1\endcsname}%
\let\auto@bib@innerbib\@empty
\bibitem [{\citenamefont {Gross}\ \emph {et~al.}(1985)\citenamefont {Gross},
  \citenamefont {Kanter},\ and\ \citenamefont {Sompolinsky}}]{gross:85}%
  \BibitemOpen
  \bibfield  {author} {\bibinfo {author} {\bibfnamefont {D.~J.}\ \bibnamefont
  {Gross}}, \bibinfo {author} {\bibfnamefont {I.}~\bibnamefont {Kanter}}, \
  and\ \bibinfo {author} {\bibfnamefont {H.}~\bibnamefont {Sompolinsky}},\
  }\bibfield  {title} {\enquote {\bibinfo {title} {{Mean-field theory of the
  Potts glass}},}\ }\href {\doibase 10.1103/PhysRevLett.55.304} {\bibfield
  {journal} {\bibinfo  {journal} {Phys. Rev. Lett.}\ }\textbf {\bibinfo
  {volume} {55}},\ \bibinfo {pages} {304--307} (\bibinfo {year}
  {1985})}\BibitemShut {NoStop}%
\bibitem [{\citenamefont {Gardner}(1985)}]{gardner:85}%
  \BibitemOpen
  \bibfield  {author} {\bibinfo {author} {\bibfnamefont {E.}~\bibnamefont
  {Gardner}},\ }\bibfield  {title} {\enquote {\bibinfo {title} {Spin glasses
  with p-spin interactions},}\ }\href {\doibase
  https://urldefense.com/v3/__https://doi.org/10.1016/0550-3213(85)90374-8__;!!PDiH4ENfjr2_Jw!C1-zBenYmQpb7Gi0Bt4QBrTS6bFF4JeEvFrbJ04msE-0JSlQ-yDIIEa8Lh1kllMoL2EfX7mnbPAjqLPZL4DBvnFkfYup$
  [doi[.]org]} {\bibfield  {journal} {\bibinfo  {journal} {Nuclear Physics B}\
  }\textbf {\bibinfo {volume} {257}},\ \bibinfo {pages} {747--765} (\bibinfo
  {year} {1985})}\BibitemShut {NoStop}%
\bibitem [{\citenamefont {Kirkpatrick}\ \emph {et~al.}(1989)\citenamefont
  {Kirkpatrick}, \citenamefont {Thirumalai},\ and\ \citenamefont
  {Wolynes}}]{kirkpatrick:89}%
  \BibitemOpen
  \bibfield  {author} {\bibinfo {author} {\bibfnamefont {T.~R.}\ \bibnamefont
  {Kirkpatrick}}, \bibinfo {author} {\bibfnamefont {D.}~\bibnamefont
  {Thirumalai}}, \ and\ \bibinfo {author} {\bibfnamefont {P.~G.}\ \bibnamefont
  {Wolynes}},\ }\bibfield  {title} {\enquote {\bibinfo {title} {{Scaling
  concepts for the dynamics of viscous liquids near an ideal glassy state}},}\
  }\href {\doibase 10.1103/PhysRevA.40.1045} {\bibfield  {journal} {\bibinfo
  {journal} {Phys. Rev. A}\ }\textbf {\bibinfo {volume} {40}},\ \bibinfo
  {pages} {1045--1054} (\bibinfo {year} {1989})}\BibitemShut {NoStop}%
\bibitem [{\citenamefont {Kirkpatrick}\ and\ \citenamefont
  {Thirumalai}(2015)}]{kirkpatrick:15}%
  \BibitemOpen
  \bibfield  {author} {\bibinfo {author} {\bibfnamefont {T.~R.}\ \bibnamefont
  {Kirkpatrick}}\ and\ \bibinfo {author} {\bibfnamefont {D.}~\bibnamefont
  {Thirumalai}},\ }\bibfield  {title} {\enquote {\bibinfo {title} {{Colloquium:
  Random first order transition theory concepts in biology and physics}},}\
  }\href {\doibase 10.1103/RevModPhys.87.183} {\bibfield  {journal} {\bibinfo
  {journal} {Rev. Mod. Phys.}\ }\textbf {\bibinfo {volume} {87}},\ \bibinfo
  {pages} {183--209} (\bibinfo {year} {2015})}\BibitemShut {NoStop}%
\bibitem [{\citenamefont {Lubchenko}\ and\ \citenamefont
  {Wolynes}(2007)}]{lubchenko:07}%
  \BibitemOpen
  \bibfield  {author} {\bibinfo {author} {\bibfnamefont {Vassiliy}\
  \bibnamefont {Lubchenko}}\ and\ \bibinfo {author} {\bibfnamefont {Peter~G.}\
  \bibnamefont {Wolynes}},\ }\bibfield  {title} {\enquote {\bibinfo {title}
  {{Theory of Structural Glasses and Supercooled Liquids}},}\ }\href
  {https://urldefense.com/v3/__https://www.annualreviews.org/doi/abs/10.1146/annurev.physchem.58.032806.104653__;!!PDiH4ENfjr2_Jw!C1-zBenYmQpb7Gi0Bt4QBrTS6bFF4JeEvFrbJ04msE-0JSlQ-yDIIEa8Lh1kllMoL2EfX7mnbPAjqLPZL4DBvh_BbwtV$
  [annualreviews[.]org]} {\bibfield  {journal} {\bibinfo  {journal} {Annual
  Review of Physical Chemistry}\ }\textbf {\bibinfo {volume} {58}},\ \bibinfo
  {pages} {235--266} (\bibinfo {year} {2007})}\BibitemShut {NoStop}%
\bibitem [{\citenamefont {Cavagna}(2009)}]{cavagna:09}%
  \BibitemOpen
  \bibfield  {author} {\bibinfo {author} {\bibfnamefont {Andrea}\ \bibnamefont
  {Cavagna}},\ }\bibfield  {title} {\enquote {\bibinfo {title} {{{Supercooled
  liquids for pedestrians}}},}\ }\href
  {https://urldefense.com/v3/__https://www.sciencedirect.com/science/article/abs/pii/S0370157309001112__;!!PDiH4ENfjr2_Jw!C1-zBenYmQpb7Gi0Bt4QBrTS6bFF4JeEvFrbJ04msE-0JSlQ-yDIIEa8Lh1kllMoL2EfX7mnbPAjqLPZL4DBvhbXB8Mu$
  [sciencedirect[.]com]} {\bibfield  {journal} {\bibinfo  {journal} {Physics
  Reports}\ }\textbf {\bibinfo {volume} {476}},\ \bibinfo {pages} {51}
  (\bibinfo {year} {2009})}\BibitemShut {NoStop}%
\bibitem [{\citenamefont {Biroli}\ and\ \citenamefont
  {Bouchaud}(2010)}]{biroli:09}%
  \BibitemOpen
  \bibfield  {author} {\bibinfo {author} {\bibfnamefont {G.}~\bibnamefont
  {Biroli}}\ and\ \bibinfo {author} {\bibfnamefont {J.~P.}\ \bibnamefont
  {Bouchaud}},\ }\bibfield  {title} {\enquote {\bibinfo {title} {{ Random
  First-Order Transition Theory of Glasses: A Critical Assessment}},}\ }in\
  \href
  {https://urldefense.com/v3/__https://onlinelibrary.wiley.com/doi/book/10.1002/9781118202470__;!!PDiH4ENfjr2_Jw!C1-zBenYmQpb7Gi0Bt4QBrTS6bFF4JeEvFrbJ04msE-0JSlQ-yDIIEa8Lh1kllMoL2EfX7mnbPAjqLPZL4DBvm3bXkgf$
  [onlinelibrary[.]wiley[.]com]} {\emph {\bibinfo {booktitle} {{Structural
  Glasses and Supercooled Liquids: Theory, Experiment,and Applications}}}},\
  \bibinfo {editor} {edited by\ \bibinfo {editor} {\bibfnamefont {P.~G.}\
  \bibnamefont {Wolynes}}\ and\ \bibinfo {editor} {\bibfnamefont
  {V.}~\bibnamefont {Lubschenko}}}\ (\bibinfo  {publisher} {Wiley, Singapore},\
  \bibinfo {year} {2010})\ Chap.~\bibinfo {chapter} {2}\BibitemShut {NoStop}%
\bibitem [{\citenamefont {Franz}\ and\ \citenamefont
  {Parisi}(1999)}]{franz:99}%
  \BibitemOpen
  \bibfield  {author} {\bibinfo {author} {\bibfnamefont {S.}~\bibnamefont
  {Franz}}\ and\ \bibinfo {author} {\bibfnamefont {G.}~\bibnamefont {Parisi}},\
  }\bibfield  {title} {\enquote {\bibinfo {title} {Critical properties of a
  three-dimensional p-spin model},}\ }\href
  {https://urldefense.com/v3/__https://link.springer.com/article/10.1007/s100510050707__;!!PDiH4ENfjr2_Jw!C1-zBenYmQpb7Gi0Bt4QBrTS6bFF4JeEvFrbJ04msE-0JSlQ-yDIIEa8Lh1kllMoL2EfX7mnbPAjqLPZL4DBvrpd5ZZj$
  [link[.]springer[.]com]} {\bibfield  {journal} {\bibinfo  {journal} {The
  European Physical Journal B - Condensed Matter and Complex Systems}\ }\textbf
  {\bibinfo {volume} {8}},\ \bibinfo {pages} {417} (\bibinfo {year}
  {1999})}\BibitemShut {NoStop}%
\bibitem [{\citenamefont {Campellone}\ \emph {et~al.}(1998)\citenamefont
  {Campellone}, \citenamefont {Coluzzi},\ and\ \citenamefont
  {Parisi}}]{campellone:98}%
  \BibitemOpen
  \bibfield  {author} {\bibinfo {author} {\bibfnamefont {Matteo}\ \bibnamefont
  {Campellone}}, \bibinfo {author} {\bibfnamefont {Barbara}\ \bibnamefont
  {Coluzzi}}, \ and\ \bibinfo {author} {\bibfnamefont {Giorgio}\ \bibnamefont
  {Parisi}},\ }\bibfield  {title} {\enquote {\bibinfo {title} {{Numerical study
  of a short-range $p$-spin glass model in three dimensions}},}\ }\href
  {https://urldefense.com/v3/__https://link.aps.org/doi/10.1103/PhysRevB.58.12081__;!!PDiH4ENfjr2_Jw!C1-zBenYmQpb7Gi0Bt4QBrTS6bFF4JeEvFrbJ04msE-0JSlQ-yDIIEa8Lh1kllMoL2EfX7mnbPAjqLPZL4DBvl8yreBY$
  [link[.]aps[.]org]} {\bibfield  {journal} {\bibinfo  {journal} {Phys. Rev.
  B}\ }\textbf {\bibinfo {volume} {58}},\ \bibinfo {pages} {12081} (\bibinfo
  {year} {1998})}\BibitemShut {NoStop}%
\bibitem [{\citenamefont {Kauzmann}(1948)}]{kauzmann:48}%
  \BibitemOpen
  \bibfield  {author} {\bibinfo {author} {\bibfnamefont {W.}~\bibnamefont
  {Kauzmann}},\ }\bibfield  {title} {\enquote {\bibinfo {title} {The nature of
  the glassy state and the behavior of liquids at low temperatures},}\ }\href
  {https://urldefense.com/v3/__https://pubs.acs.org/doi/10.1021/cr60135a002__;!!PDiH4ENfjr2_Jw!C1-zBenYmQpb7Gi0Bt4QBrTS6bFF4JeEvFrbJ04msE-0JSlQ-yDIIEa8Lh1kllMoL2EfX7mnbPAjqLPZL4DBvlU0OfCQ$
  [pubs[.]acs[.]org]} {\bibfield  {journal} {\bibinfo  {journal} {Chem.
  Reviews}\ }\textbf {\bibinfo {volume} {43}},\ \bibinfo {pages} {219}
  (\bibinfo {year} {1948})}\BibitemShut {NoStop}%
\bibitem [{\citenamefont {Yeo}\ and\ \citenamefont
  {Moore}(2012{\natexlab{a}})}]{yeoB:12}%
  \BibitemOpen
  \bibfield  {author} {\bibinfo {author} {\bibfnamefont {Joonhyun}\
  \bibnamefont {Yeo}}\ and\ \bibinfo {author} {\bibfnamefont {M.~A.}\
  \bibnamefont {Moore}},\ }\bibfield  {title} {\enquote {\bibinfo {title}
  {{Renormalization group analysis of the $M$-$p$-spin glass model with $p=3$
  and $M=3$}},}\ }\href {\doibase 10.1103/PhysRevB.85.100405} {\bibfield
  {journal} {\bibinfo  {journal} {Phys. Rev. B}\ }\textbf {\bibinfo {volume}
  {85}},\ \bibinfo {pages} {100405} (\bibinfo {year}
  {2012}{\natexlab{a}})}\BibitemShut {NoStop}%
\bibitem [{\citenamefont {Yeo}\ and\ \citenamefont
  {Moore}(2012{\natexlab{b}})}]{yeo:13}%
  \BibitemOpen
  \bibfield  {author} {\bibinfo {author} {\bibfnamefont {Joonhyun}\
  \bibnamefont {Yeo}}\ and\ \bibinfo {author} {\bibfnamefont {M.~A.}\
  \bibnamefont {Moore}},\ }\bibfield  {title} {\enquote {\bibinfo {title}
  {{Origin of the growing length scale in $M$-$p$-spin glass models}},}\ }\href
  {\doibase 10.1103/PhysRevE.86.052501} {\bibfield  {journal} {\bibinfo
  {journal} {Phys. Rev. E}\ }\textbf {\bibinfo {volume} {86}},\ \bibinfo
  {pages} {052501} (\bibinfo {year} {2012}{\natexlab{b}})}\BibitemShut
  {NoStop}%
\bibitem [{\citenamefont {Yeo}\ and\ \citenamefont
  {Moore}(2020)}]{yeo2020possible}%
  \BibitemOpen
  \bibfield  {author} {\bibinfo {author} {\bibfnamefont {J.}~\bibnamefont
  {Yeo}}\ and\ \bibinfo {author} {\bibfnamefont {M.~A.}\ \bibnamefont
  {Moore}},\ }\bibfield  {title} {\enquote {\bibinfo {title} {{Possible
  instability of one-step replica symmetry breaking in $p$-spin Ising models
  outside mean-field theory}},}\ }\href {\doibase 10.1103/PhysRevE.101.032127}
  {\bibfield  {journal} {\bibinfo  {journal} {Phys. Rev. E}\ }\textbf {\bibinfo
  {volume} {101}},\ \bibinfo {pages} {032127} (\bibinfo {year}
  {2020})}\BibitemShut {NoStop}%
\bibitem [{\citenamefont {Moore}(2006)}]{moore:06b}%
  \BibitemOpen
  \bibfield  {author} {\bibinfo {author} {\bibfnamefont {M.~A.}\ \bibnamefont
  {Moore}},\ }\bibfield  {title} {\enquote {\bibinfo {title} {{Interface Free
  Energies in $p$-Spin Glass Models}},}\ }\href
  {https://urldefense.com/v3/__https://journals.aps.org/prl/abstract/10.1103/PhysRevLett.96.137202__;!!PDiH4ENfjr2_Jw!C1-zBenYmQpb7Gi0Bt4QBrTS6bFF4JeEvFrbJ04msE-0JSlQ-yDIIEa8Lh1kllMoL2EfX7mnbPAjqLPZL4DBvtVnayZF$
  [journals[.]aps[.]org]} {\bibfield  {journal} {\bibinfo  {journal} {Phys.
  Rev. Lett.}\ }\textbf {\bibinfo {volume} {96}},\ \bibinfo {pages} {137202}
  (\bibinfo {year} {2006})}\BibitemShut {NoStop}%
\bibitem [{\citenamefont {Rudnick}\ and\ \citenamefont
  {Nelson}(1976)}]{rudnick:76}%
  \BibitemOpen
  \bibfield  {author} {\bibinfo {author} {\bibfnamefont {Joseph}\ \bibnamefont
  {Rudnick}}\ and\ \bibinfo {author} {\bibfnamefont {David~R.}\ \bibnamefont
  {Nelson}},\ }\bibfield  {title} {\enquote {\bibinfo {title} {{Equations of
  state and renormalization-group recursion relations}},}\ }\href {\doibase
  10.1103/PhysRevB.13.2208} {\bibfield  {journal} {\bibinfo  {journal} {Phys.
  Rev. B}\ }\textbf {\bibinfo {volume} {13}},\ \bibinfo {pages} {2208--2221}
  (\bibinfo {year} {1976})}\BibitemShut {NoStop}%
\bibitem [{\citenamefont {H\"oller}\ and\ \citenamefont
  {Read}(2020)}]{holler:20}%
  \BibitemOpen
  \bibfield  {author} {\bibinfo {author} {\bibfnamefont {J.}~\bibnamefont
  {H\"oller}}\ and\ \bibinfo {author} {\bibfnamefont {N.}~\bibnamefont
  {Read}},\ }\bibfield  {title} {\enquote {\bibinfo {title} {{One-step
  replica-symmetry-breaking phase below the de Almeida--Thouless line in
  low-dimensional spin glasses}},}\ }\href {\doibase
  10.1103/PhysRevE.101.042114} {\bibfield  {journal} {\bibinfo  {journal}
  {Phys. Rev. E}\ }\textbf {\bibinfo {volume} {101}},\ \bibinfo {pages}
  {042114} (\bibinfo {year} {2020})}\BibitemShut {NoStop}%
\bibitem [{\citenamefont {de~Almeida}\ and\ \citenamefont
  {Thouless}(1978)}]{almeida:78}%
  \BibitemOpen
  \bibfield  {author} {\bibinfo {author} {\bibfnamefont {J~R~L}\ \bibnamefont
  {de~Almeida}}\ and\ \bibinfo {author} {\bibfnamefont {D~J}\ \bibnamefont
  {Thouless}},\ }\bibfield  {title} {\enquote {\bibinfo {title} {Stability of
  the sherrington-kirkpatrick solution of a spin glass model},}\ }\href
  {\doibase 10.1088/0305-4470/11/5/028} {\bibfield  {journal} {\bibinfo
  {journal} {Journal of Physics A: Mathematical and General}\ }\textbf
  {\bibinfo {volume} {11}},\ \bibinfo {pages} {983} (\bibinfo {year}
  {1978})}\BibitemShut {NoStop}%
\bibitem [{\citenamefont {Aguilar-Janita}\ \emph {et~al.}(2023)\citenamefont
  {Aguilar-Janita}, \citenamefont {Martin-Mayor}, \citenamefont
  {Moreno-Gordo},\ and\ \citenamefont {Ruiz-Lorenzo}}]{aguilar:23}%
  \BibitemOpen
  \bibfield  {author} {\bibinfo {author} {\bibfnamefont {Miguel}\ \bibnamefont
  {Aguilar-Janita}}, \bibinfo {author} {\bibfnamefont {Victor}\ \bibnamefont
  {Martin-Mayor}}, \bibinfo {author} {\bibfnamefont {Javier}\ \bibnamefont
  {Moreno-Gordo}}, \ and\ \bibinfo {author} {\bibfnamefont {Juan~Jesus}\
  \bibnamefont {Ruiz-Lorenzo}},\ }\href@noop {} {\enquote {\bibinfo {title}
  {{Second order phase transition in the six-dimensional Ising spin glass on a
  field}},}\ } (\bibinfo {year} {2023}),\ \Eprint
  {http://arxiv.org/abs/2306.00569} {arXiv:2306.00569 [cond-mat.dis-nn]}
  \BibitemShut {NoStop}%
\bibitem [{\citenamefont {Caltagirone}\ \emph {et~al.}(2011)\citenamefont
  {Caltagirone}, \citenamefont {Ferrari}, \citenamefont {Leuzzi}, \citenamefont
  {Parisi},\ and\ \citenamefont {Rizzo}}]{caltagirone:11}%
  \BibitemOpen
  \bibfield  {author} {\bibinfo {author} {\bibfnamefont {F.}~\bibnamefont
  {Caltagirone}}, \bibinfo {author} {\bibfnamefont {U.}~\bibnamefont
  {Ferrari}}, \bibinfo {author} {\bibfnamefont {L.}~\bibnamefont {Leuzzi}},
  \bibinfo {author} {\bibfnamefont {G.}~\bibnamefont {Parisi}}, \ and\ \bibinfo
  {author} {\bibfnamefont {T.}~\bibnamefont {Rizzo}},\ }\bibfield  {title}
  {\enquote {\bibinfo {title} {{Ising $M$-$p$-spin mean-field model for the
  structural glass: Continuous versus discontinuous transition}},}\ }\href
  {\doibase 10.1103/PhysRevB.83.104202} {\bibfield  {journal} {\bibinfo
  {journal} {Phys. Rev. B}\ }\textbf {\bibinfo {volume} {83}},\ \bibinfo
  {pages} {104202} (\bibinfo {year} {2011})}\BibitemShut {NoStop}%
\bibitem [{\citenamefont {Aspelmeier}\ \emph {et~al.}(2008)\citenamefont
  {Aspelmeier}, \citenamefont {Billoire}, \citenamefont {Marinari},\ and\
  \citenamefont {Moore}}]{aspelmeier:08}%
  \BibitemOpen
  \bibfield  {author} {\bibinfo {author} {\bibfnamefont {T}~\bibnamefont
  {Aspelmeier}}, \bibinfo {author} {\bibfnamefont {A}~\bibnamefont {Billoire}},
  \bibinfo {author} {\bibfnamefont {E}~\bibnamefont {Marinari}}, \ and\
  \bibinfo {author} {\bibfnamefont {M~A}\ \bibnamefont {Moore}},\ }\bibfield
  {title} {\enquote {\bibinfo {title} {{Finite-size corrections in the
  Sherrington-Kirkpatrick model}},}\ }\href {\doibase
  10.1088/1751-8113/41/32/324008} {\bibfield  {journal} {\bibinfo  {journal}
  {Journal of Physics A: Mathematical and Theoretical}\ }\textbf {\bibinfo
  {volume} {41}},\ \bibinfo {pages} {324008} (\bibinfo {year}
  {2008})}\BibitemShut {NoStop}%
\bibitem [{\citenamefont {Goldbart}\ and\ \citenamefont
  {Elderfield}(1985)}]{goldbart:85}%
  \BibitemOpen
  \bibfield  {author} {\bibinfo {author} {\bibfnamefont {P}~\bibnamefont
  {Goldbart}}\ and\ \bibinfo {author} {\bibfnamefont {D}~\bibnamefont
  {Elderfield}},\ }\bibfield  {title} {\enquote {\bibinfo {title} {{The failure
  of the Parisi scheme for spin glass models without reflection symmetry}},}\
  }\href {\doibase 10.1088/0022-3719/18/9/009} {\bibfield  {journal} {\bibinfo
  {journal} {Journal of Physics C: Solid State Physics}\ }\textbf {\bibinfo
  {volume} {18}},\ \bibinfo {pages} {L229} (\bibinfo {year}
  {1985})}\BibitemShut {NoStop}%
\bibitem [{\citenamefont {Jani\ifmmode~\check{s}\else \v{s}\fi{}}\ \emph
  {et~al.}(2013)\citenamefont {Jani\ifmmode~\check{s}\else \v{s}\fi{}},
  \citenamefont {Kauch},\ and\ \citenamefont {Kl\'{\i}\ifmmode~\check{c}\else
  \v{c}\fi{}}}]{janis:13}%
  \BibitemOpen
  \bibfield  {author} {\bibinfo {author} {\bibfnamefont {V.}~\bibnamefont
  {Jani\ifmmode~\check{s}\else \v{s}\fi{}}}, \bibinfo {author} {\bibfnamefont
  {A.}~\bibnamefont {Kauch}}, \ and\ \bibinfo {author} {\bibfnamefont
  {A.}~\bibnamefont {Kl\'{\i}\ifmmode~\check{c}\else \v{c}\fi{}}},\ }\bibfield
  {title} {\enquote {\bibinfo {title} {{Free energy of mean-field spin-glass
  models: Evolution operator and perturbation expansion}},}\ }\href {\doibase
  10.1103/PhysRevB.87.054201} {\bibfield  {journal} {\bibinfo  {journal} {Phys.
  Rev. B}\ }\textbf {\bibinfo {volume} {87}},\ \bibinfo {pages} {054201}
  (\bibinfo {year} {2013})}\BibitemShut {NoStop}%
\bibitem [{\citenamefont {Parisi}(1980)}]{Parisi_1980}%
  \BibitemOpen
  \bibfield  {author} {\bibinfo {author} {\bibfnamefont {G}~\bibnamefont
  {Parisi}},\ }\bibfield  {title} {\enquote {\bibinfo {title} {{Magnetic
  properties of spin glasses in a new mean field theory}},}\ }\href {\doibase
  10.1088/0305-4470/13/5/047} {\bibfield  {journal} {\bibinfo  {journal}
  {Journal of Physics A: Mathematical and General}\ }\textbf {\bibinfo {volume}
  {13}},\ \bibinfo {pages} {1887} (\bibinfo {year} {1980})}\BibitemShut
  {NoStop}%
\bibitem [{\citenamefont {Bray}\ and\ \citenamefont {Roberts}(1980)}]{bray:80}%
  \BibitemOpen
  \bibfield  {author} {\bibinfo {author} {\bibfnamefont {A~J}\ \bibnamefont
  {Bray}}\ and\ \bibinfo {author} {\bibfnamefont {S~A}\ \bibnamefont
  {Roberts}},\ }\bibfield  {title} {\enquote {\bibinfo {title}
  {{Renormalisation-group approach to the spin glass transition in finite
  magnetic fields}},}\ }\href {\doibase 10.1088/0022-3719/13/29/019} {\bibfield
   {journal} {\bibinfo  {journal} {Journal of Physics C: Solid State Physics}\
  }\textbf {\bibinfo {volume} {13}},\ \bibinfo {pages} {5405} (\bibinfo {year}
  {1980})}\BibitemShut {NoStop}%
\bibitem [{\citenamefont {Angelini}\ and\ \citenamefont
  {Biroli}(2015)}]{angelinibiroli:15}%
  \BibitemOpen
  \bibfield  {author} {\bibinfo {author} {\bibfnamefont {Maria~Chiara}\
  \bibnamefont {Angelini}}\ and\ \bibinfo {author} {\bibfnamefont {Giulio}\
  \bibnamefont {Biroli}},\ }\bibfield  {title} {\enquote {\bibinfo {title}
  {{Spin Glass in a Field: A New Zero-Temperature Fixed Point in Finite
  Dimensions}},}\ }\href {\doibase 10.1103/PhysRevLett.114.095701} {\bibfield
  {journal} {\bibinfo  {journal} {Phys. Rev. Lett.}\ }\textbf {\bibinfo
  {volume} {114}},\ \bibinfo {pages} {095701} (\bibinfo {year}
  {2015})}\BibitemShut {NoStop}%
\bibitem [{\citenamefont {Parisi}\ \emph {et~al.}(2014)\citenamefont {Parisi},
  \citenamefont {Ricci-Tersenghi},\ and\ \citenamefont {Rizzo}}]{rizzo:14}%
  \BibitemOpen
  \bibfield  {author} {\bibinfo {author} {\bibfnamefont {G}~\bibnamefont
  {Parisi}}, \bibinfo {author} {\bibfnamefont {F}~\bibnamefont
  {Ricci-Tersenghi}}, \ and\ \bibinfo {author} {\bibfnamefont {T}~\bibnamefont
  {Rizzo}},\ }\bibfield  {title} {\enquote {\bibinfo {title} {{Diluted
  mean-field spin-glass models at criticality}},}\ }\href {\doibase
  10.1088/1742-5468/2014/04/P04013} {\bibfield  {journal} {\bibinfo  {journal}
  {Journal of Statistical Mechanics: Theory and Experiment}\ }\textbf {\bibinfo
  {volume} {2014}},\ \bibinfo {pages} {P04013} (\bibinfo {year}
  {2014})}\BibitemShut {NoStop}%
\bibitem [{\citenamefont {Vedula}\ \emph {et~al.}(2023)\citenamefont {Vedula},
  \citenamefont {Moore},\ and\ \citenamefont {Sharma}}]{bharadwaj:23}%
  \BibitemOpen
  \bibfield  {author} {\bibinfo {author} {\bibfnamefont {Bharadwaj}\
  \bibnamefont {Vedula}}, \bibinfo {author} {\bibfnamefont {M.~A.}\
  \bibnamefont {Moore}}, \ and\ \bibinfo {author} {\bibfnamefont {Auditya}\
  \bibnamefont {Sharma}},\ }\bibfield  {title} {\enquote {\bibinfo {title}
  {{Study of the de Almeida--Thouless line in the one-dimensional diluted
  power-law $XY$ spin glass}},}\ }\href {\doibase 10.1103/PhysRevE.108.014116}
  {\bibfield  {journal} {\bibinfo  {journal} {Phys. Rev. E}\ }\textbf {\bibinfo
  {volume} {108}},\ \bibinfo {pages} {014116} (\bibinfo {year}
  {2023})}\BibitemShut {NoStop}%
\bibitem [{\citenamefont {Moore}(2012)}]{moore:12}%
  \BibitemOpen
  \bibfield  {author} {\bibinfo {author} {\bibfnamefont {M.~A.}\ \bibnamefont
  {Moore}},\ }\bibfield  {title} {\enquote {\bibinfo {title} {{$1/m$ expansion
  in spin glasses and the de Almeida-Thouless line}},}\ }\href {\doibase
  10.1103/PhysRevE.86.031114} {\bibfield  {journal} {\bibinfo  {journal} {Phys.
  Rev. E}\ }\textbf {\bibinfo {volume} {86}},\ \bibinfo {pages} {031114}
  (\bibinfo {year} {2012})}\BibitemShut {NoStop}%
\bibitem [{\citenamefont {McMillan}(1984)}]{mcmillan:84}%
  \BibitemOpen
  \bibfield  {author} {\bibinfo {author} {\bibfnamefont {W.~L.}\ \bibnamefont
  {McMillan}},\ }\bibfield  {title} {\enquote {\bibinfo {title} {{Domain-wall
  renormalization-group study of the three-dimensional random Ising model}},}\
  }\href {\doibase 10.1103/PhysRevB.30.476} {\bibfield  {journal} {\bibinfo
  {journal} {Phys. Rev. B}\ }\textbf {\bibinfo {volume} {30}},\ \bibinfo
  {pages} {476--477} (\bibinfo {year} {1984})}\BibitemShut {NoStop}%
\bibitem [{\citenamefont {Bray}\ and\ \citenamefont {Moore}(1986)}]{bray:86}%
  \BibitemOpen
  \bibfield  {author} {\bibinfo {author} {\bibfnamefont {A.~J.}\ \bibnamefont
  {Bray}}\ and\ \bibinfo {author} {\bibfnamefont {M.~A.}\ \bibnamefont
  {Moore}},\ }\bibfield  {title} {\enquote {\bibinfo {title} {Scaling theory of
  the ordered phase of spin glasses},}\ }in\ \href
  {https://urldefense.com/v3/__https://doi.org/10.1007/BFb0057515__;!!PDiH4ENfjr2_Jw!C1-zBenYmQpb7Gi0Bt4QBrTS6bFF4JeEvFrbJ04msE-0JSlQ-yDIIEa8Lh1kllMoL2EfX7mnbPAjqLPZL4DBvmm6duwH$
  [doi[.]org]} {\emph {\bibinfo {booktitle} {{Heidelberg Colloquium on Glassy
  Dynamics and Optimization}}}},\ \bibinfo {editor} {edited by\ \bibinfo
  {editor} {\bibfnamefont {L.}~\bibnamefont {Van~Hemmen}}\ and\ \bibinfo
  {editor} {\bibfnamefont {I.}~\bibnamefont {Morgenstern}}}\ (\bibinfo
  {publisher} {Springer},\ \bibinfo {address} {New York},\ \bibinfo {year}
  {1986})\ p.\ \bibinfo {pages} {121}\BibitemShut {NoStop}%
\bibitem [{\citenamefont {Fisher}\ and\ \citenamefont
  {Huse}(1988)}]{fisher:88}%
  \BibitemOpen
  \bibfield  {author} {\bibinfo {author} {\bibfnamefont {Daniel~S.}\
  \bibnamefont {Fisher}}\ and\ \bibinfo {author} {\bibfnamefont {David~A.}\
  \bibnamefont {Huse}},\ }\bibfield  {title} {\enquote {\bibinfo {title}
  {Equilibrium behavior of the spin-glass ordered phase},}\ }\href {\doibase
  10.1103/PhysRevB.38.386} {\bibfield  {journal} {\bibinfo  {journal} {Phys.
  Rev. B}\ }\textbf {\bibinfo {volume} {38}},\ \bibinfo {pages} {386--411}
  (\bibinfo {year} {1988})}\BibitemShut {NoStop}%
\bibitem [{\citenamefont {Imry}\ and\ \citenamefont {Ma}(1975)}]{imry:75}%
  \BibitemOpen
  \bibfield  {author} {\bibinfo {author} {\bibfnamefont {Yoseph}\ \bibnamefont
  {Imry}}\ and\ \bibinfo {author} {\bibfnamefont {Shang-keng}\ \bibnamefont
  {Ma}},\ }\bibfield  {title} {\enquote {\bibinfo {title} {{Random-Field
  Instability of the Ordered State of Continuous Symmetry}},}\ }\href {\doibase
  10.1103/PhysRevLett.35.1399} {\bibfield  {journal} {\bibinfo  {journal}
  {Phys. Rev. Lett.}\ }\textbf {\bibinfo {volume} {35}},\ \bibinfo {pages}
  {1399--1401} (\bibinfo {year} {1975})}\BibitemShut {NoStop}%
\bibitem [{\citenamefont {Moore}\ and\ \citenamefont {Yeo}(2006)}]{moore:06}%
  \BibitemOpen
  \bibfield  {author} {\bibinfo {author} {\bibfnamefont {M.~A.}\ \bibnamefont
  {Moore}}\ and\ \bibinfo {author} {\bibfnamefont {J.}~\bibnamefont {Yeo}},\
  }\bibfield  {title} {\enquote {\bibinfo {title} {{Thermodynamic Glass
  Transition in Finite Dimensions}},}\ }\href {\doibase
  10.1103/PhysRevLett.96.095701} {\bibfield  {journal} {\bibinfo  {journal}
  {Phys. Rev. Lett.}\ }\textbf {\bibinfo {volume} {96}},\ \bibinfo {pages}
  {095701} (\bibinfo {year} {2006})}\BibitemShut {NoStop}%
\bibitem [{\citenamefont {Larson}\ \emph {et~al.}(2010)\citenamefont {Larson},
  \citenamefont {Katzgraber}, \citenamefont {Moore},\ and\ \citenamefont
  {Young}}]{larson:10}%
  \BibitemOpen
  \bibfield  {author} {\bibinfo {author} {\bibfnamefont {Derek}\ \bibnamefont
  {Larson}}, \bibinfo {author} {\bibfnamefont {Helmut~G.}\ \bibnamefont
  {Katzgraber}}, \bibinfo {author} {\bibfnamefont {M.~A.}\ \bibnamefont
  {Moore}}, \ and\ \bibinfo {author} {\bibfnamefont {A.~P.}\ \bibnamefont
  {Young}},\ }\bibfield  {title} {\enquote {\bibinfo {title} {{Numerical
  studies of a one-dimensional three-spin spin-glass model with long-range
  interactions}},}\ }\href {\doibase 10.1103/PhysRevB.81.064415} {\bibfield
  {journal} {\bibinfo  {journal} {Phys. Rev. B}\ }\textbf {\bibinfo {volume}
  {81}},\ \bibinfo {pages} {064415} (\bibinfo {year} {2010})}\BibitemShut
  {NoStop}%
\end{thebibliography}%
\end{document}